\documentclass[letterpaper,twocolumn,showpacs,preprintnumbers,amsmath,amssymb,floatfix,superscriptaddress,pre]{revtex4}

\usepackage{psfrag,color}
\usepackage{graphicx}  
\usepackage{dcolumn}   
\usepackage{bm}        
\usepackage{epsfig}
\usepackage{textcomp}
\usepackage[sort&compress]{natbib}

\begin{document}

\title{Modulation and correlation lengths in
  systems with 
competing interactions}

\author{Saurish Chakrabarty}
\affiliation{Department of Physics and Center for Materials Innovation, Washington University in St Louis, MO 63130.}

\author{Zohar Nussinov}
\affiliation{Department of Physics and Center for Materials Innovation, Washington University in St Louis, MO 63130.}
\affiliation{Kavli Institute for Theoretical Physics, Santa Barbara, CA 93106.}

\date{\today}

\begin{abstract}
We examine correlation functions in the presence of competing long and
short ranged interactions to find multiple correlation and modulation
lengths.
We calculate the ground state stripe width of an Ising ferromagnet, frustrated by an arbitrary long range interaction.
In large $n$ systems, we demonstrate that
for a short range system frustrated by a general competing long range
interaction, the crossover temperature 
$T^*$ veers towards the critical temperature of the unfrustrated short
range system (i.e., that in which the frustrating long
range interaction is removed) .
We also show that apart from certain special crossover points,
the total number of correlation and modulation lengths remains
conserved. We derive an expression
for the change in modulation length with temperature for a general
system near the ground state
with a ferromagnetic interaction and an opposing long range
interaction. We illustrate that the correlation functions associated
with the exact dipolar interactions differ substantially from those in which a
scalar product form between the dipoles is assumed.
\end{abstract}


\maketitle

\section{Introduction}
Short range interactions have been at the focus of much
study for many decades. Perhaps one of the best known examples are the Ising ferromagnet and the anti-ferromagnet \cite{ising}. Long range interactions are equally abundant \cite{long}. 
Systems in which both long and short range interactions co-exist
comprise very interesting systems. Such competing forces can lead to a
wealth of interesting patterns -- stripes, bubbles, etc. \cite{lieb},
\cite{vindigni}, \cite{ortix}, \cite{gulacsi}, \cite{barci}.   Realizations are found in numerous fields -- quantum Hall systems \cite{Fogler}, adatoms on metallic surfaces, amphiphilic systems \cite{amp}, interacting elastic defects (dislocations and
disclinations) in solids \cite{vortex3}, interactions amongst vortices
in fluid mechanics 
\cite{vortex1} and superconductors \cite{vortex2}, crumpled membrane
systems \cite{Seul}, wave-particle interactions \cite{wavpart},
interactions amongst holes in cuprate superconductors 
\cite{steve}, \cite{us}, \cite{zohar}, \cite{Low}, \cite{carlson}, arsenide superconductors \cite{nasci}, manganates and nickelates 
\cite{cheong}, \cite{golosov}, some theories of structural glasses \cite{dk}, \cite{peter}, 
\cite{Gilles}, \cite{new}, \cite{reviewGilles}, colloidal systems \cite{der},
\cite{reich} and many many more. 
Much of the work to date focused
on the character of the transitions in these systems and the subtle
thermodynamics that is often observed (e.g., the equivalence between
different ensembles in many such systems is no longer as obvious, nor
always correct, as it is in the canonical short range case
\cite{barre}). Other very interesting aspects of different systems
have been addressed in \cite{azbel}. 

Here we investigate the general temperature dependence
of the structural features that appear in such systems when competing
interactions of short and long range are present.
Our focus here is on large $n$ spin systems.
In many such systems, 
there are emergent modulation lengths governing the size of various domains. 
We find that {\em these modulation lengths often adhere to various
  scaling laws, sharp crossovers and divergences at various
  temperatures} (with no associated thermodynamic transition). We also
find that in such systems, correlation lengths generically evolve into
modulation lengths (and vice versa) at various temperatures. The
behavior of correlation and modulation lengths as a function of
temperature will afford us with certain selection rules on the
possible underlying microscopic interactions. 
In their simplest incarnation, our central results are as follows:

\begin{enumerate}
\item
In canonical systems harboring competing short (finite) and
long range interactions modulated patterns appear. Depending on the
type of the long range interaction, 
the modulation length either increases or decreases from its ground
state value as the temperature is raised.
We will relate this change, in lattice systems, 
to derivatives of the Fourier transforms
of the interactions that are present.  
\item
There exist special crossover temperatures at which new
correlation/modulation lengths come up or some cease to exist.
The total number
of characteristic length-scales (correlation + modulation) remains
conserved, except at the crossover points.
\item
The presence of the angular dependent dipolar interaction term
that frustrates an otherwise unfrustrated ferromagnet vis
a vis a simple scalar product between the dipoles adds 
new (dominant) length-scales. The angular dependence
significantly changes the system.
\end{enumerate}

We will further investigate the ground state modulation
lengths in general frustrated Ising systems and also
point to discontinuous jumps in the modulation
lengths that may appear in the large $n$ rendition
of some systems. 

\bigskip

\begin{figure}[ht]
  \begin{center}
    \includegraphics[width=\columnwidth]{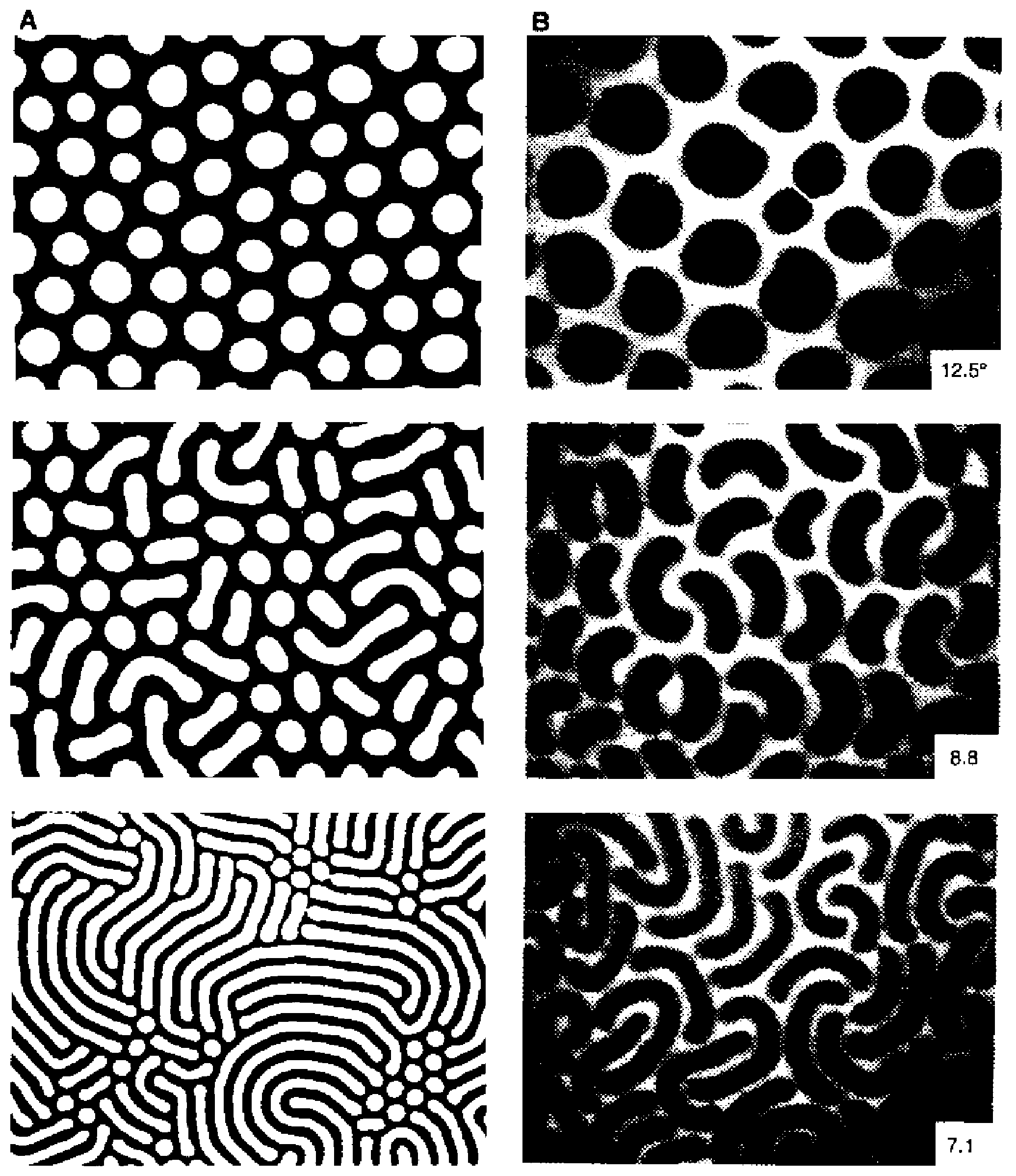}
    \caption{Reproduced with permission from Science, Ref.\cite{Seul}. Reversible ``strip-out'' instability in magnetic and organic thin films. Period ($L_{D}$) reduction under the constraint of fixed overall composition and fixed number of domains leads to elongation of bubbles. Left panel (A) in magnetic garnet films, this is achieved by raising the temperature [labeled in (B) in degrees Celsius] along the symmetry axis, $H=0$ (period in bottom panel, $\sim 10 \mu m$) (see Fig. 5). Right panel (B) In Langmuir films composed of phospholipid dimyristolphosphatidic acid (DMPA) and cholesterol (98:2 molar ratio, pH 11), this is achieved by lowering the temperature at constant average molecular density [period in bottom panel, $\sim 20 \mu m$]}. 
    \label{andelman}
  \end{center}
\end{figure}

Armed with these general results, we may discern the viable
microscopic interactions (exact or effective) 
which underlie temperature dependent patterns that are triggered by two competing
interactions. Our 
analysis suggests the effective microscopic interactions that may drive non-uniform
patterns such as those underlying lattice analogs of the systems
of Fig. \ref{andelman}. 

The treatment that we present in this work applies to lattice
systems and does not account for the curvature of bubbles and other
continuum objects. These may be augmented by inspecting energy
functionals (and their associated free energy extrema) of various 
continuum field morphologies under the addition of detailed domain
wall tension forms -- e.g., explicit line 
integrals along the perimeter where surface tension exists -- and the
imposition of additional constraints via Lagrange multipliers. 
We leave their analysis for future work. One of the central results
of our work is the derivation of conditions relating to the
increase/decrease
of modulation lengths in lattice systems with changes in temperature. These 
conditions relate the change in the modulation length at
low temperatures to the derivatives of
the Fourier transforms of the interactions present.

In Section(\ref{systems_of_study}), we outline the general systems that
we study. In particular, we introduce the frustrated ferromagnet.

In Section(\ref{Isingg}), we derive the scaling
form for the Ising ground states for general
frustrating long range interactions. 
Henceforth, we provide explicit expressions for
the crossover temperatures and the correlations 
lengths in the large $n$ limit.

In Section(\ref{largeN}), we introduce the two spin correlation
function for a general system in this limit.

Based on the correlation function, we then present
some general results for systems with competing nearest neighbor
ferromagnetic interaction and an arbitrary long range interaction in
Section(\ref{results}). We start by deriving the equilibrium stripe
width for a two dimensional Ising system with nearest neighbor
ferromagnetic interactions and competing long range interactions.
We derive an expression for the change in
modulation length with temperature for low temperatures for large $n$ systems. We illustrate
how the crossover temperature,
$T^*$ arises in the large $n$ limit and show some general properties of the system associated with it.

We present some example systems in Section(\ref{appn}). We
numerically calculate the correlation function for the screened
Coulomb frustrated ferromagnet and the dipolar frustrated
ferromagnet. We then study the screened Coulomb frustrated ferromagnet
in more details. Next, we show some results for systems with higher
dimensional spins.  We study a system with the dipole-dipole interaction for
three dimensional spins, without ignoring the angular dependent term and
show that this term changes the ground state length-scales of the
system considerably. We also present a system with the
Dzyaloshinsky - Moriya interaction in addition to the ferromagnetic
term and a general frustrating long range term.

We give our concluding remarks in Section(\ref{conc}).

\section{The systems of study}
\label{systems_of_study}

Consider a translationally invariant system whose Hamiltonian is given by 
\begin{eqnarray}
H = \frac{1}{2} \sum_{\vec{x}\neq\vec{y}}V(|\vec{x}-\vec{y}|)S(\vec{x}) S(\vec{y}). 
\label{Ham}
\end{eqnarray} 
The quantities $\{S(\vec{x})\}$ portray classical scalar spins or
fields. The sites $\vec{x}$ and $\vec{y}$ lie on a hypercubic lattice
with $N$ sites having unit lattice constant. In what follows, $v(k)$
and $s(\vec{k})$ will denote the Fourier transforms of
$V(|\vec{x}-\vec{y}|)$ and $S(\vec{x})$.
Thus, we have,
\begin{eqnarray}
H=\frac{1}{2N}\sum_{\vec{k}}v(k)|s(\vec{k})|^2
\end{eqnarray}
For analytic interactions,
$v(k)$ is a function of $k^{2}$ (to avoid branch cuts).

The two point correlator function for the system in Eq. (\ref{Ham})
is,
\begin{eqnarray}
G(\vec{x})=\langle S(0)S(\vec{x}) \rangle
\end{eqnarray}
Thus, in the Fourier space, we have,
\begin{eqnarray}
G(\vec{k})=\langle |s(\vec{k})|^2 \rangle
\end{eqnarray}
Throughout most of this work we will
focus on systems with competing interactions
having Hamiltonians of the following form.
\begin{equation}
H=-J\sum_{ \langle \vec{x},\vec{y} \rangle}S(\vec{x})S(\vec{y})+Q\sum_{\vec{x} \neq \vec{y}}V_L(|\vec{x} - \vec{y}|)S(\vec{x})S(\vec{y}),
\label{Ham_JQ}
\end{equation}
where the first term represents nearest neighbor ferromagnetic
interaction for positive $J$ and the second term represents some long
range interaction which opposes the ferromagnetic interaction for positive $Q$.
We will study properties of general systems of the form of Eq.(\ref{Ham_JQ}). 
In order to flesh out the physical meaning of our results 
and illustrate their implications and meaning, 
we will further provide explicit expressions and numerical results for 
two particular examples.
The Hamiltonian of Eq.(\ref{Ham_JQ}) represents a system that we christen to be the screened {\em Coulomb Frustrated Ferromagnet} when
\begin{eqnarray}
V_L(r)=\frac{e^{-\lambda r}}{4\pi r} \mbox{ in three dimensions, and} \nonumber\\
V_L(r)=K_0(\lambda r) \mbox{ in two dimensions,}
\label{ccs}
\end{eqnarray}
where $\lambda^{-1}$ represents the screening length
and $K_{0}$ is a modified Bessel function,
\begin{eqnarray}
K_{0}(x) = \int_{0}^{\infty} dt \frac{\cos xt}{\sqrt{1+t^{2}}}.
\label{K_0}
\end{eqnarray}
Throughout our work, we will discuss both the screened and unscreened renditions 
of the Coulomb frustrated ferromagnet.
Eq.(\ref{Ham_JQ}) corresponds to a {\em Dipolar Frustrated Ferromagnet} when
\begin{eqnarray}
V_L(r)&=&\frac{1}{r^3}\nonumber\\
&=&\frac{1}{(r^2+\delta^2)^{3/2}}\mbox{ in the limit }\delta\to 0,
\label{dipint}
\end{eqnarray}
on the lattices that we will consider. 
Later, we also consider the general direction dependent 
(relative to the location vectors) of the dipolar
interaction for three dimensional spins; we will replace
the scalar product form of the dipolar interactions in Eqs.(\ref{Ham_JQ}, \ref{dipint}) 
by the precise dipolar interactions
between magnetic moments.

On a hypercubic lattice, the nearest neighbor interactions in real space of Eq.(\ref{Ham_JQ})
have the lattice Laplacian
\begin{eqnarray} 
\Delta(\vec{k}) = 2 \sum_{l=1}^{d} (1-\cos k_{l}) ,
\label{Laplace0}
\end{eqnarray}
as their Fourier transform. In the continuum (small $k$) limit, $\Delta \to z = k^{2}$. The real lattice Laplacian
\begin{equation}
\langle \vec{x}| \Delta |\vec{y} \rangle = \left\{ \begin{array}{ll}
2d & \mbox{ for $\vec{x}=\vec{y}$} \\
-1 & \mbox{ for $|\vec{x}-\vec{y}| = 1$.}
\end{array}
\right.
\label{Laplace1}
\end{equation}
Notice that  $\langle \vec{x}| \Delta^{R} |\vec{y} \rangle = 0 
\mbox{ for $ |\vec{x}-\vec{y} |> R$}$,
where $R$ is the spatial range over which the interaction kernel
is non-vanishing. The following corresponds to interactions
of Range=2 lattice constants, 
\begin{eqnarray}
  \langle \vec{x}| \Delta^{2}| \vec{y} \rangle \mbox{ } 
= \mbox{ }&& 2d(2d+2) \mbox{
    for } \vec{x} = \vec{y} \nonumber
  \\
  \ \ && -4d \mbox{ for  } |\vec{x}-\vec{y}| = 1 \nonumber \\
  \ \ && 2 \mbox{ for } (\vec{x} -\vec{y}) = 
(\pm \hat{e}_{\ell} \pm \hat{e}_{\ell^{\prime}}) \mbox{ where  }
 \ell \neq \ell^{\prime}      \nonumber \\
  \ \ && 1 \mbox{ for a $ \pm 2 \hat{e}_{\ell} $ separation}.
\label{Laplace2}
\end{eqnarray}

Correspondingly, in the continuum, the Lattice Laplacian and its powers
attain simple forms and capture tendencies in numerous systems.
Surface tension in many systems is captured by a $g(\nabla \phi)^{2}$ term where $\phi$ is a constant in a uniform domain. Upon Fourier transforming, such squared gradient terms lead to a $k^{2}$ 
dependence. The effects of curvature which are notable in many mixtures and membrane systems
can often be emulated by terms involving $(\nabla^{2} h)$ with $h$ a variable parameterizing the profile; at times the interplay of such curvature terms with others leads, in the aftermath, to a simple short range $k^{4}$ term in the interaction kernel (the continuum version of the squared
lattice Laplacian of Eq.(\ref{Laplace2})). An excellent review of
 these issues is addressed in \cite{Seul}.

\section{Ground state stripe width for Ising systems}
\label{Isingg}
We next briefly discuss the ground
state stripe width for systems with
general long range interactions in Eq.(\ref{Ham_JQ}). Below, we discuss the 
Ising ground states. We will later on consider the spherical model
that will enable us to compute the correlation functions at arbitrary temperatures.
We consider a system with Ising spins in $d$ dimension and assume that the system
forms a ``striped'' pattern (periodic pattern along one of the
dimensions -- stripes in two dimensions, parallel slices in three dimensions 
and so on) of spin-up and spin-down states of period
$l$. We then calculate the free energy as a function of $l$ and then
minimize it with respect to $l$ to get the equilibrium stripe width.
For the frustrated ferromagnet, if $v_L(k)=1/k^p$,
\begin{eqnarray}
l=\left[\frac{(2\pi)^{p+2}J}{4Qp\left(1-\frac{1}{2^{p+2}}\right)\zeta(p+2)}\right]^\frac{1}{p+1},
\end{eqnarray}
where $\zeta(s)$ is the Riemann zeta function,
\begin{eqnarray}
\zeta(s)=\sum_{n=1}^\infty n^{-s}.
\end{eqnarray}
For the particular case of Coulomb frustration, $p=2$,
\begin{eqnarray}
l=\sqrt[3]{\frac{64\pi^2}{5}\frac{J}{Q}},
\end{eqnarray}
in accord with the results of Refs. \cite{Gilles, zohar_com}. 
For long range dipolar interactions ($p=3$), we find that
\begin{eqnarray}
l=\delta\sqrt{\frac{3J}{Q}}.
\end{eqnarray}
In the notation to be employed later on in this work, $l$ 
plays the role of the modulation length of the system
at zero temperature, $L_{D}(T=0)$.

\section{Correlation Functions in the large $n$ limit - general considerations}
\label{largeN}
The results reported henceforth were computed within the spherical or large $n$ limit \cite{kac}. 
It was found by Stanley long ago \cite{stanley} that the 
large $n$ limit of the $n$ component normalized spin systems (so called $O(n)$ spins)
is identical
to the spherical model first introduced by Berlin and Kac \cite{kac}.

The designation of ``$O(n)$ spins''
simply denotes real fields (spins) of unit length that
have some arbitrary number $(n)$
of components. For $n=1$, the system
is an Ising model. A single component real field
having unit norm allows for only two scalars at any given site $\vec{x}$: $S(\vec{x}) = \pm 1$.
The $n=2$ system corresponds to a two component spin system in which
the spins are free to rotate in a two dimensional place -- $S_{1}^{2}(\vec{x}) + S_{2}^{2}(\vec{x}) =1$
(the so-called XY spin system).
The case of $n=3$ corresponds to a system
of  three component Heisenberg type spins, and so on. 
In general,
\begin{eqnarray}
\sum_{a=1}^{n} S_{a}(\vec{x}) S_{a}(\vec{x})=1.
\end{eqnarray}

We now
introduce the spherical model in its generality. The spins in Eq.(\ref{Ham})
satisfy a single global (``spherical'') constraint,
\begin{eqnarray}
\sum_{\vec{x}} \langle S^{2}(\vec{x}) \rangle = N,
\label{con}
\end{eqnarray}
enforced by a Lagrange multiplier $\mu$.
 This leads to the functional $H'=H+\mu N$ which renders the model quadratic (as both Eqs.(\ref{Ham}, \ref{con}) are quadratic) and thus exactly solvable, see, e.g., \cite{us}. 
The continuum analogs of Eqs.(\ref{Ham}, \ref{con})
read
\begin{eqnarray}
H = \frac{1}{2} \int d^{d}x~ d^{d}y~ V(|\vec{x}-\vec{y}|) S(\vec{x}) S(\vec{y}), \nonumber
\\ \int d^{d} x ~\langle S^{2}(\vec{x}) \rangle = const.
\end{eqnarray} 
The quadratic theory may be solved exactly. From the equipartition
theorem, for $T\ge T_c$, the Fourier space correlator 
\begin{equation}
G(k)=\langle |S^2(k)|\rangle=\frac{k_BT}{v(k)+\mu}.
\label{sksq}
\end{equation}
The real space two point correlator is given by
\begin{eqnarray}
G(\vec{x}) \equiv \langle S(0) S( \vec{x}) \rangle = k_{B} T
\int_{BZ} \frac{d^{d}k}{(2 \pi)^{d}} \frac{e^{i\vec{k} \cdot \vec{x}}}
{v(k) + \mu},
\label{corr}
\end{eqnarray}
with $d$ the spatial dimension and $BZ$ denoting the integration over
the first Brillouin zone. For a hypercubic lattice in $d$
dimensions with a lattice constant that is set to one,
 $-\pi < k_{i} \le \pi$ for $i=1,2, ..., d$. Henceforth,  to avoid cumbersome notation, 
 we will generally drop the designation of $BZ$; this is to be understood on 
 all momentum space integrals pertaining to the lattice
 systems that we examine. To complete the characterization of
the correlation functions at different temperatures, we note that the
Lagrange multiplier $\mu(T)$ is given by the implicit equation $1=
G(\vec{x}=0)$.
Thus,
\begin{equation}
1=k_BT\int\frac{d^dk}{(2\pi)^d}\frac{1}{v(k)+\mu}.
\label{g1mut}
\end{equation}
This implies that the temperature $T$ is a monotonic increasing function of $\mu$. If
$\mu$ changes by a small amount $\Delta\mu$,  then $T$ will change by an amount $\Delta
T$, such that 
\begin{equation}
\Delta T\propto\Delta\mu.
\label{deltdelmu}
\end{equation}
Eq.(\ref{g1mut}) also implies that in the high temperature limit,
\begin{eqnarray}
\frac{\mu}{k_BT}&=&\frac{\pi^{d/2}\Lambda^d}{(2\pi)^d\Gamma(\frac{d}{2}+1)}\\
\implies T&\propto&\mu,
\label{tpropmu}
\end{eqnarray}
where $\Lambda$ is the upper limit of the $k$ integration,
representing the ultra-violet cut-off in continuum renditions of 
the large $n$ system.  Similarly, if we perform, for a lattice system, 
the momentum
integration 
 in a hypercube of side $2\pi$, we have in the high
temperature limit,
\begin{eqnarray}
\mu=k_BT.
\end{eqnarray}
Furthermore, as the integration range in Eq. (\ref{g1mut})
  is finite, we can prove that $\mu(T)$ is an analytic
function of $T$ [See Appendix \ref{mutanalytic}]. This supports the assumption that $G^{-1}(T,k)$ is
analytic in $T$ and $k$ at all points except $k=0$ where $v(k)$ is
usually singular.

We investigate the general character of the correlation functions given
by Eq.(\ref{corr}) for rotationally invariant systems. If the minimum
(minima) of $v(k)$ occur(s) at momenta $q$ far from the Brillouin zone
boundaries of the cubic lattice then we may set the range of
integration in Eq.(\ref{corr}) to be unrestricted. The correlation
function is then dominated by the location of the poles (and/or branch
cuts) of $1/[v(k)+\mu]$.
Thus, we look for solutions to the following equation.
\begin{equation}
v(k)+\mu=0.
\label{poleseq1}
\end{equation}
The system is perfectly ordered in its ground state. From a
temperature at which the system is not perfectly ordered, as we lower
the temperature, the correlation length diverges at $T=T_c$.
At $T=T_c$,
$\mu$ takes the value,
\begin{equation}
\mu_{min}=-\min_{k\in BZ}[v(k)].
\label{mumin}
\end{equation}
As the temperature is increased, the disorder creeps in and in many
systems, at a temperature $T^*$, the modulation length diverges.

The characteristic length-scales of the system are governed by the poles of $[v(k)+\mu]^{-1}$.
\begin{equation}
J\Delta(\vec{k})+Qv_L(k)+\mu=0,
\end{equation}
which in the continuum limit takes the form
\begin{equation}
Jk^2+Qv_L(k)+\mu=0.
\label{poleseq}
\end{equation}
Employing the above considerations, we will derive, in the next section, some general results for
systems  of the form
Eq.(\ref{Ham_JQ}).

Our work will focus on classical systems. The extension to the quantum arena \cite{us}
 is straightforward. 
In, e.g., large $n$ bosonic renditions of our system,  replicating the usual large $n$ analysis,
we find \cite{new} that the pair correlator
\begin{eqnarray}
G_{B}(\vec{k}) = \frac{n_{B}\left(\sqrt{\frac{v(\vec{k}) + \mu}{k_{B} T}}\right) + \frac{1}{2}}{\sqrt{\frac{v(\vec{k}) + \mu}{k_{B} T}}},
\label{gbose}
\end{eqnarray}
with the bosonic distribution function
\begin{eqnarray}
n_B(x) =\frac{1}{e^x-1}.
\end{eqnarray}

The correlator of Eq.(\ref{gbose}) is of a similar nature as that
of the classical correlator of Eq.(\ref{sksq}) with branch cuts
generally appearing in the quantum case. Our analysis below
relies on the evolution of the poles of $v(k) + \mu$ as a function
of temperature in classical systems.

In the quantum arena, we first choose the proper contour in the
complex $k$-space (going around the branch cuts).
Then, we argue that the only points that contribute to the integral
are the points where the integrand is singular.
This corresponds to $v(k)+\mu=0$. 
Thus, the integral remains unchanged if we expand the integrand to lowest
order in $v(k)+\mu$. Doing this, we get, to leading order,
\begin{eqnarray}
G_B(k)=\frac{k_BT}{v(k)+\mu}
\end{eqnarray}
which is same as the classical expression.
The characteristic length-scales of the system are therefore still determined by
the zeros of $v(k)+\mu$ in the complex $k$ plane in the exact same way.

For interactions that are not isotropic, for both classical and
quantum cases, we need to consider the full six dimensional space of
the complex components of $\vec{k}$ along each of the three coordinate axes.

\section{Large $n$ Results}
\label{results}

In this section, we present some general results for systems of the form
Eq.(\ref{Ham_JQ}) in their large $n$ limit. First, we find the dependence of the modulation
length on temperature, near $T_c$. Next, we will illustrate an analogy between the behavior of the
correlation length near the critical temperature $T_c$ and that of the
modulation length near $T^*$.
We will then discuss some aspects of the crossover points.
We end this section with some results in the
high temperature limit.

\subsection{The low temperature limit: a criterion for determining an increase
or decrease of the modulation length at low temperatures}
\label{ltl}
In this section, we derive universal conditions for increasing or decreasing
modulation lengths in general systems pairwise interactions.
Eqs.(\ref{selruleodd}, \ref{selruleeven}) show general conditions for the change in
modulation length, $L_D$ with temperature for a general system of the form
Eq.(\ref{Ham_JQ}).
The value, $k_0$ of $k$ which satisfies Eq.(\ref{mumin}),
\begin{eqnarray}
v(k_{0}) = \min_{k \in BZ} v(\vec{k})
\label{kmin}
\end{eqnarray}
determines the modulation length at $T=T_c$.
\begin{eqnarray}
v(k_0)+\mu_{min}=0,\\
v'(k_0)=0.
\label{k_0}
\end{eqnarray}
As the temperature is raised, the new pole near $k_0$ will have an imaginary part corresponding to the finite correlation length. The real part will also change in general and this would induce a change in the modulation length. Let $\mu(T)=\mu_{min}+\delta\mu$ with $\delta\mu>0$. Then we have,
\begin{eqnarray}
k=k_0+\delta k,\nonumber\\
\delta k=\sum_{j=1}^{\infty}\delta k_j,
\end{eqnarray}
where $\delta k_j\propto\delta\mu^{x_j}$, $x_{j+1}>x_j$. Our goal is
to find the leading order real contribution to $\delta k$ which would
give us the change in modulation length with increasing $\mu$ and hence with increasing temperature.
\begin{equation}
\sum_{j=2}^\infty v^{(j)}(k_0)\frac{\delta k^j}{j!}+\delta\mu=0.
\end{equation}
Suppose $v^{(n)}(k_0)=0$ for $2<n<p$ and $v^{(p)}(k_0)\neq 0$.
[Clearly, in  most cases, the third derivative is not zero and $p=3$.]
We have,
\begin{eqnarray}
\frac{v^{(2)}(k_0)}{2!})(\delta k_1^2+2\delta k_1\delta k_2 + ... )\nonumber\\
+ [\frac{v^{(p)}(k_0)}{p!}(\delta k_1^p+p\delta k_1^{p-1}\delta k_2+ ...)
+ \frac{v^{(p+1)}(k_0)}{(p+1)!}\times\nonumber\\(\delta k_1^{p+1}+(p+1)\delta k_1^{p}\delta k_2+ ...) + ... ] +\delta\mu=0.\ \ 
\end{eqnarray}
To leading order,
\begin{eqnarray}
\frac{v^{(2)}(k_0)}{2!}\delta k_1^2+\delta\mu=0,\nonumber\\
\delta k_1^2=-\frac{2\delta\mu}{v^{(2)}(k_0)}.
\label{k1corr}
\end{eqnarray}
From this, we see that $\delta k_1$ is imaginary. This constitutes another verification of
the well established result about the universality of the divergence
of the correlation length, $\xi$ at $T_c$ with the mean-field type 
critical exponent $\nu =1/2$ in the large $n$ limit. 
\begin{eqnarray}
\xi\propto(T-T_c)^{-\nu},\nonumber\\
\nu=\frac{1}{2}.
\label{nutc}
\end{eqnarray}
The next, higher order, relations are obtained using the method of dominant balance.
\begin{eqnarray}
v^{(2)}(k_0) (\delta k_1)(\delta k_2)+\frac{v^{(p)}(k_0)}{p!}(\delta k_1)^p=0\nonumber\\
\delta k_2=\frac{(-1)^{\frac{p+1}{2}}v^{(p)}(k_0)(\delta\mu)^{\frac{p-1}{2}}}{2 p!(\frac{v^{(2)}(k_0)}{2!})^{\frac{p+1}{2}}}.
\end{eqnarray}
Therefore, $\delta k_2$ is real if $p$ is odd and imaginary if $p$ is even. 
If,
\begin{eqnarray}
L_D(T)=L_D(T_c)+\delta L_D,
\end{eqnarray}
then, for $p=2n+1$, 
\begin{equation}
\delta L_D=\frac{2\pi}{k_0^2}\frac{(-1)^nv^{(2n+1)}(k_0)\delta\mu^n}{2(2n+1)!(\frac{v^{(2)}(k_0)}{2!})^{n+1}}.
\label{selruleodd}
\end{equation}
Thus to get the leading order real contribution to $\delta k$ for even $p$$[>2]$, we have to go to higher order.
\begin{eqnarray}
2(\frac{v^{(2)}(k_0)}{2!})\delta k_1\delta k_3+\frac{v^{(p+1)}(k_0)}{(p+1)!}\delta k_1^{p+1}=0\nonumber\\
\delta k_3=\frac{(-1)^{1+p/2}v^{(p+1)}(k_0)(\delta\mu)^{p/2}}{2(p+1)!(\frac{v^{(2)}(k_0)}{2!})^{p/2+1}}.
\end{eqnarray}
For $p=2n$,
\begin{equation}
\delta L_D=\frac{2\pi}{k_0^2}\frac{(-1)^{n}v^{(2n+1)}(k_0)(\delta\mu)^{n}}{2(2n+1)!(\frac{v^{(2)}(k_0)}{2!})^{n+1}}.
\label{selruleeven}
\end{equation}
If, for $p=2n$, $v^{(2n+1)}(k_0)=0$, then we will need to continue this process until we get a real contribution to $\delta k$. In appendix \ref{explicit_form_p}, we provide explicit forms for $\delta L_{D}$ for different values of
$p$.

In the most common case, where $v^{(3)}(k_0)\neq0$, we have,
\begin{eqnarray}
\delta L_D=-\frac{2\pi}{k_0^2}\frac{v^{(3)}(k_0)}{3[v^{(2)}(k_0)]^2}\delta\mu.
\end{eqnarray}
Also, applying this to a nearest neighbor system in the continuum frustrated 
by a general long range interaction given by $v_L(k)$ in Fourier space , we get,
\begin{eqnarray}
\delta L_D=-\frac{2\pi}{k_0^2}\frac{Qv_L^{(3)}(k_0)}{3[v_L^{(2)}(k_0)]^2}\delta\mu.
\end{eqnarray}
This shows that it is the long range term that determines the sign of the change in modulation length with temperature
as the system is heated from $T=T_C$.
{\em The results derived above allow us to relate an
  increase/decrease in the modulation length at low temperatures
to the sign of the first non-vanishing odd derivative (of an order larger
than two) of the Fourier transform of the interactions 
that are present.} It is important to emphasize that our results
apply to any interaction. These may include screened or unscreened Coulomb
and other long range interactions but may also include interactions
that are strictly of finite range [e.g., next-nearest neighbor
interactions
on the lattice for which we have $v_{L}  = -t \Delta^{2}$ (with a constant
$t>0$, see Eq.(\ref{Laplace2}))].

The results from this section about modulation lengths just above $T_C$, can give us
similar behavior of the correlation lengths at temperatures slightly below $T^*$.

\subsection{A correspondence between the temperature $T^*$ at which the
  modulation length diverges and the critical temperature $T_c$}
The critical temperature $T_c$ corresponds to the maximum value of $\mu$ for which
Eq.(\ref{poleseq1}) still attains a real solution. Thus,
\begin{eqnarray}
v(k_0)+\mu_{min}=0,\nonumber\\
v'(k_0)=0,\nonumber\\
v''(k_0)>0.
\label{ktc}
\end{eqnarray}
For systems in which the modulation length diverges at $T^*$, $T^*$ corresponds to the minimum value of $\mu$ for
which Eq.(\ref{poleseq1}) has a purely imaginary solution. Thus, if $v(i\kappa)=\hat{v}(\kappa)$,
\begin{eqnarray}
\hat{v}(\kappa_0)+\mu^*=0\nonumber\\
\hat{v}'(\kappa_0)=0\nonumber\\
\hat{v}''(\kappa_0)<0\implies v''(i\kappa_0)>0
\label{kappatstar}
\end{eqnarray}
Thus, we expect similar qualitative results for the correlation lengths
at temperatures slightly above $T_c$ as for modulation lengths
slightly below $T^*$ and vice-versa. [The relations for the
  derivatives of $\hat{v}(\kappa_0)$ in Eq.(\ref{kappatstar}) are
  guaranteed to hold only if $T^*>T_c$.]

\subsection{Crossover temperatures: Emergent modulations}
For systems with competing multiple range interactions, there
may exist special temperatures at which the poles of the correlation
function change character, thus changing modulation lengths to
correlation lengths and vice-versa.
In particular, for most systems we have a crossover temperature $T^*$
above which the system does not have any modulation.
Apart from this kind of phenomenon, there might also be finite
discontinuous jumps in the modulation length. This is illustrated with
an example in Section(\ref{1storder}).

\begin{figure}[h]
  \begin{center}
    \includegraphics[width=\columnwidth]{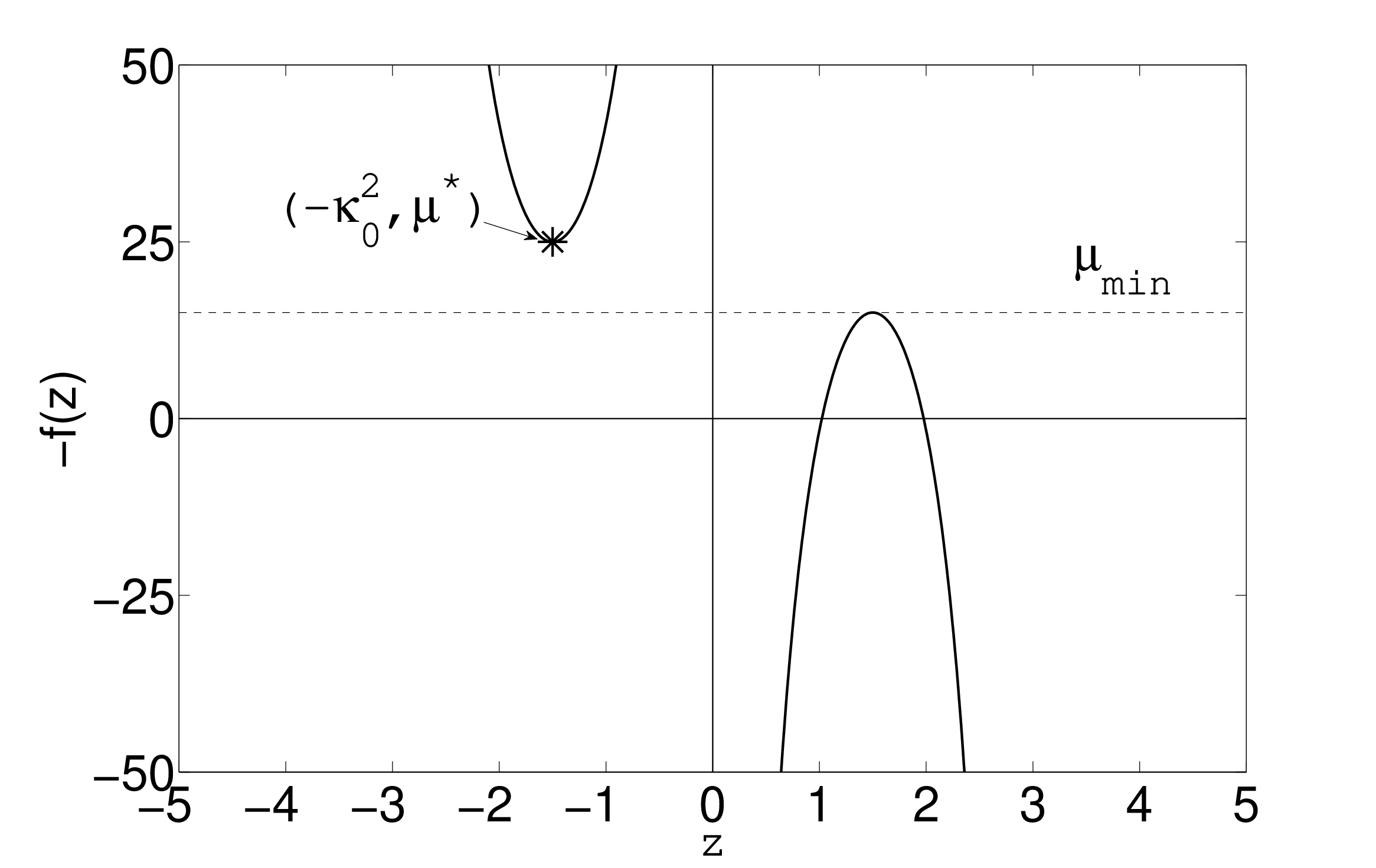}
    \caption{$-f(z)=-v(k)=\mu$ plotted against $z=k^2$ for purely real and
    purely imaginary $k$'s ($T\to0$ and $T\to\infty$
    respectively). The negative $z$ regime corresponds to temperatures
    (Lagrange multiplier, $\mu$'s) for which purely imaginary solutions
    exist. The maximum of the curve in the positive $z$ regime corresponds to the
    modulations at $T=T_c$ [$\mu=\mu_{min}$], which is the maximum temperature at which
    pure modulations exist.}
    \label{fzmumin}
  \end{center}
\end{figure}

\begin{figure}[ht]
  \begin{center}
    \epsfig{file=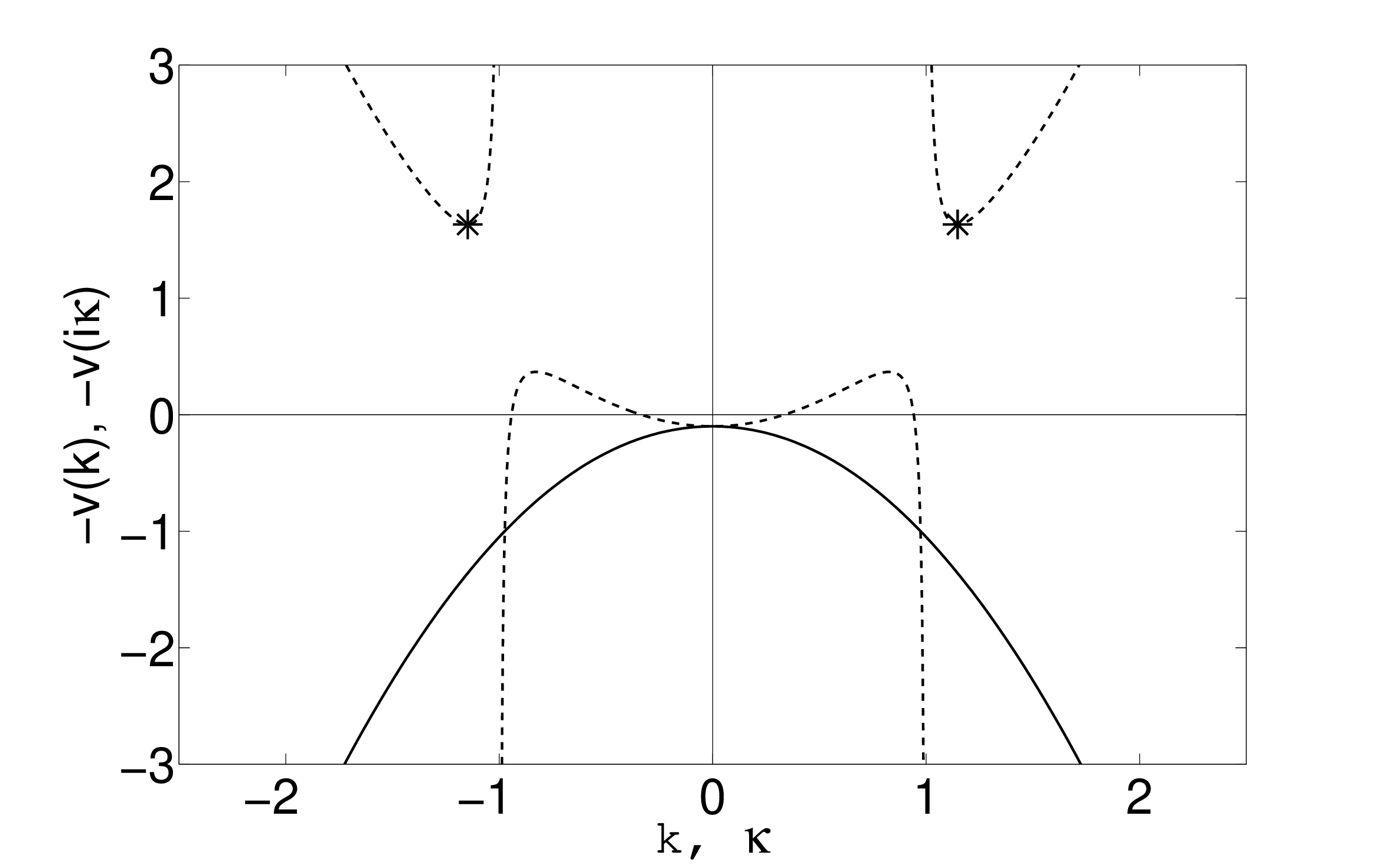,width=\columnwidth}
    \epsfig{file=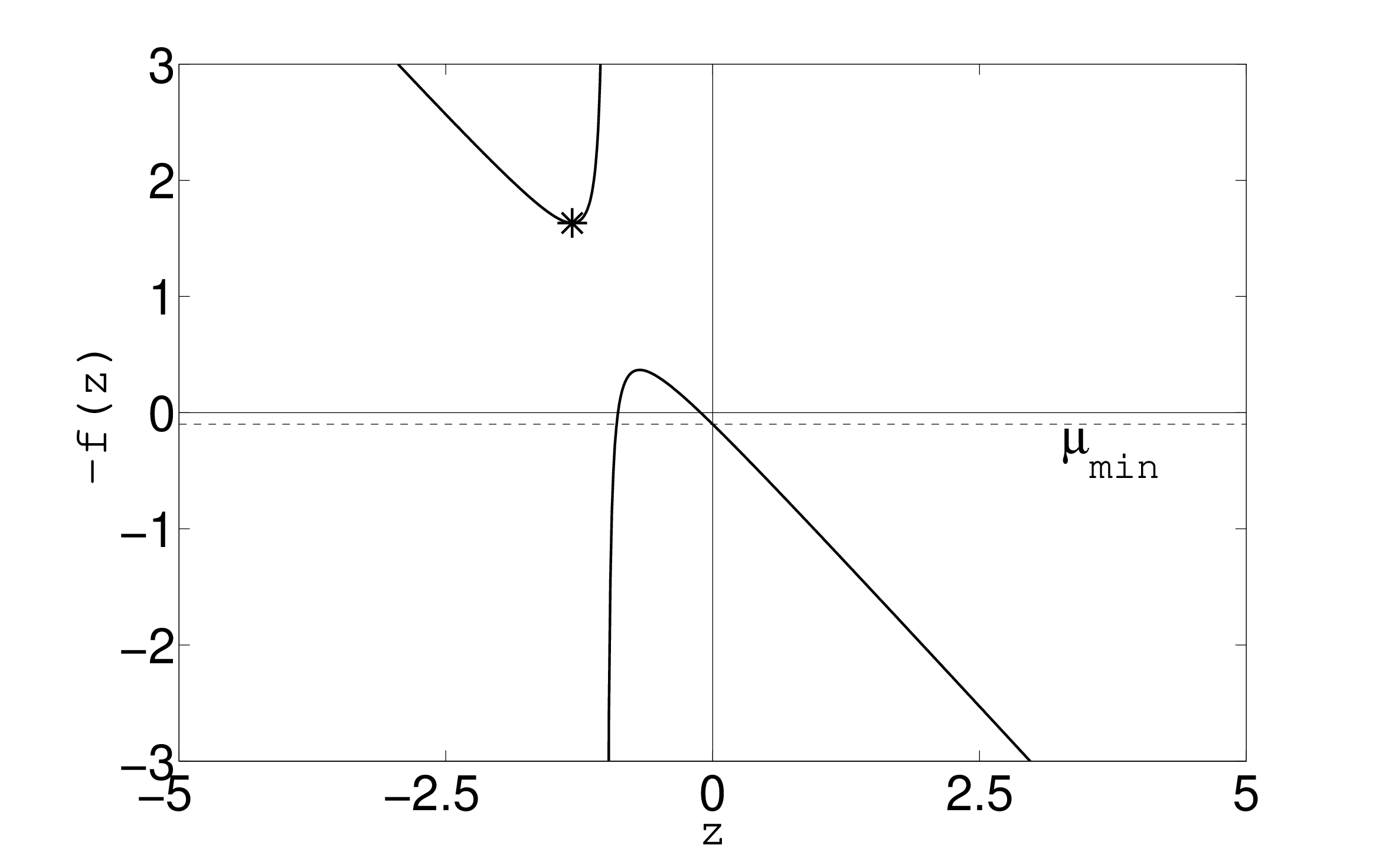,width=\columnwidth}
    \caption{TOP: Solid line:$-v(k)$ plotted against $k$.\\
      Dashed line:$-v(i\kappa)$ plotted against $\kappa$.\\
      BOTTOM: $-f(z)$ plotted against $z$.}
    \label{mustar_phys}
  \end{center}
\end{figure}

We start by defining the crossover temperature $T^*$ for a
ferromagnetic system frustrated by a general long range interaction.
Let $k=i\kappa$, $\kappa\in\mathbb{R}$ above $T^*$ and
$\kappa=\kappa_0$ at $T^*$.
Let $v(k)=f(z)$, $z=k^2$. Above $T^*$,
$\mu=\mu_{min}+\Delta\mu$ [$\Delta\mu>0$]. Using Eq.(\ref{mumin}),
\begin{eqnarray}
\mu=-f(-\kappa^2)&=&-\min_{k\in\mathbb{R}}v(k)+\Delta\mu\nonumber\\
&=&\max_{k\in\mathbb{R}}[-v(k)]+\Delta\mu.
\end{eqnarray}
$T^*$ corresponds to the minimum value of $\Delta\mu$ for which we
have at least one such solution. See Fig. \ref{fzmumin}.
Thus,
\begin{eqnarray}
\mu^*&=&\min_{\begin{array}{c}\kappa\in\mathbb{R}\\-v(i\kappa)\ge\mu_{min}\end{array}}[-v(i\kappa)]\nonumber\\
&=&-\max_{\begin{array}{c}\kappa\in\mathbb{R}\\-v(i\kappa)\ge\mu_{min}\end{array}}[v(i\kappa)].
\label{mustar}
\end{eqnarray}
Sometimes, the crossover point is slightly more difficult to
visualize. See Fig. \ref{mustar_phys}. In this case, the minimum
upper branch of $-f(z)$ for $z<0$ [equivalently the upper branches of
$-v(i\kappa)$] gives us the value of $\mu^*$. The branch chosen has to
continue to $\mu=+\infty$ so that at least one term without modulation is
always available as we increase the temperature, as required by the
definition of $T^*$. The other branch provides such solutions only up to
a certain temperature. Also, the part of it which is below $\mu_{min}$
is irrelevant.

If $f(z)$ is an odd function of $z$ (e.g. the Coulomb frustrated
ferromagnet), $\mu^*=-\mu_{min}$ and the correlation length at $T^*$
is the same as the modulation length at $T_c$.

Also, for the system in Eq.(\ref{Ham_JQ}), if
$\lim_{\kappa\to0}v_L(i\kappa)=+\infty$, we have, $\mu^*=\mu_{min}$, and $T^*=T_c$.

\subsubsection{$T^*=T_c$ if all the competing interactions
  are of finite range and crossover exists}
For systems where all the competing interactions are of finite range,
$T^*=T_c$. We prove this as follows.
Since finite range interactions contribute to $v(k)$ as powers of
$\Delta(k)\to k^2$, for a
system with only finite range interactions, $f(z)$ is analytic for all
$z$. For a minimum of $-f(z)$ to exist in the $z<0$ regime which is higher than
the maximum in the $z>0$ regime, we need $f(z)$ to be discontinuous at
some point. Putting all of the pieces together, we find that there are
no possibilities:  (i) no crossover, i.e., $T^*=\infty$ or (ii) $\kappa_0=0$ 
and $\mu^*=\mu_{min}$ with $T^*=T_c$.

\subsubsection{$T^*\to T_c$ as the strength of the long range
  interaction is turned off}
The results from this section and the next hold for a general system,
not just the frustrated ferromagnet.

The crossover temperature
$T^*$ tends to $T_c$ for $Q=0$ as $Q\to0$.
For a general system, let $G(T,k)$ denote the Fourier space
correlation function at temperature $T$.
By definition, at $T=T_c$ the correlation length is infinite. Thus,
$T_c$ is the solution to
\begin{eqnarray}
G^{-1}(T,k)=0,
\label{Ginv}
\end{eqnarray}
such that $k \in BZ$ (or for continuum renditions,
$k\in\mathbb{R}$).

$T^*$ is the temperature at which the modulation length
diverges for the frustrated ferromagnet, or becomes the same as the
modulation length of the unfrustrated system at $T_c$ for a general system. Thus, $T^*$ is the solution to
\begin{eqnarray}
G^{-1}(T,q+i\kappa)=0,
\label{Ginv_im}
\end{eqnarray}
with $\kappa\in\mathbb{R}$ [$q=0$ for the case of the frustrated
  ferromagnet, $q=\pi$ for the frustrated anti-ferromagnet].
At $T_c$, for $Q=0$, we have,
\begin{eqnarray}
G^{-1}(T_c,q)=0.
\end{eqnarray}
This however also satisfies Eq.(\ref{Ginv_im}), with $\kappa=0$. Therefore,
\begin{eqnarray}
\lim_{Q\to0}T^*=T_c.
\end{eqnarray}

\begin{figure}[hb]
  \begin{center}
    \includegraphics[width=\columnwidth]{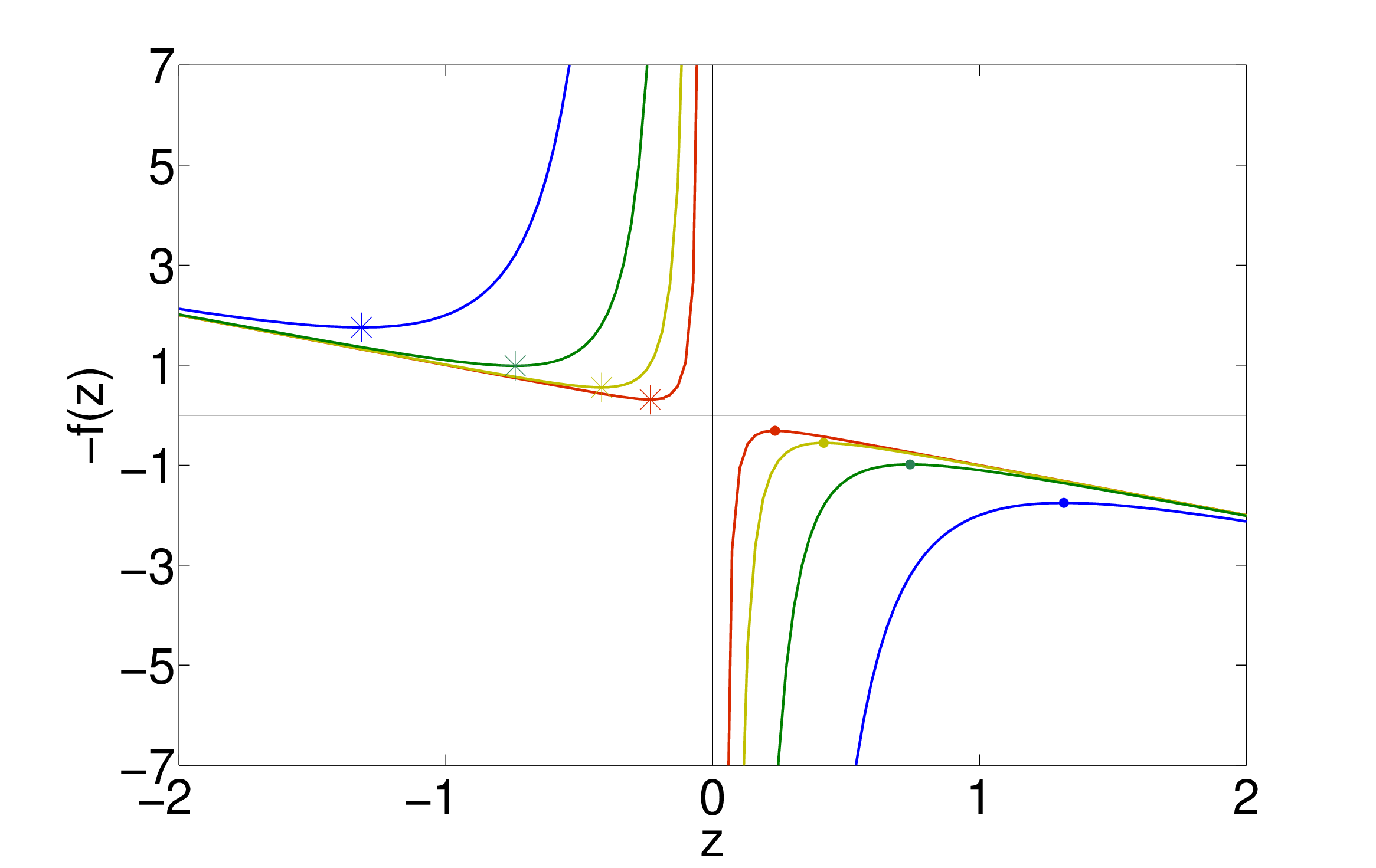}
    \caption{(Color online.) Illustration of the limit $T^*\to T_c$ as $Q\to0$ with
      $v_L(k)=1/k^3$. The plot shows $-f(z)=-v(k)$ vs $z=k^2$, for $v(k)=Jk^2+Q/k^3$ with $J=1$ and
      $Q=\{1-$Blue, $0.1-$Green, $0.01-$Yellow, $0.001-$Red$\}$. `*'
      represents the value of $\mu^*$ and dot represents $\mu_{min}$.}
    \label{qto0}
  \end{center}
\end{figure}
We demonstrate this in the large $n$ limit. See Fig. \ref{qto0}.
For $Q=0$, we have
$\mu_{min}=0$ and $k_0=0$. Let $v_L(k)$ diverge as
$k^{-2p}$ near $k=0$. For small $Q$, from Eq.(\ref{mumin}), we have, 
\begin{eqnarray}
k_0=(\frac{pQ}{J})^\frac{1}{2p+2},\nonumber\\
\mu_{min}=-\frac{p+1}{p^{\frac{p}{p+1}}}J^{\frac{p}{p+1}}Q^{\frac{1}{p+1}}.
\end{eqnarray}
If $p$ is odd,
\begin{eqnarray}
\mu^*=-\mu_{min},\nonumber\\
\kappa_0=k_0.
\end{eqnarray}
As $Q\to 0$, $\kappa_0=k_0=0$ and $\mu^*=\mu_{min}=0$, that is,
\begin{equation}
\lim_{Q\to0}T^*=T_c(Q=0).
\end{equation}

\subsubsection{Proof of the conservation of the total number of characteristic
  length-scales}
  \label{conser}
In this section we consider the general situation
in which the Fourier transform of the interaction
kernel, $v(k)$, is a rational function of $z=\Delta(k)$, [$z\to k^2$ in the
continuum limit]. That is, we consider
situations for which 
\begin{equation}
v(k)=f(z)=\frac{P(z)}{Q(z)},
\label{PQ}
\end{equation}
with
$P$ and $Q$ polynomials (of degrees $M_1$ and $M_2$ respectively).
We will now demonstrate that the combined sum of the number of correlation and the number of
modulation lengths remains unchanged as the temperature is varied.

Before providing the proof of our assertion, we first re-iterate
that the form of Eq.(\ref{PQ}) is rather general. 
For a system with finite range interactions
($V(|\vec{x}- \vec{y}| >R) =0$ with finite $R$)
that is even under parity ($V(\vec{x}- \vec{y}) = V(\vec{y}- \vec{x})$),
the Fourier transform of $V(\vec{x}-\vec{y})$ can be written as a finite order
polynomial
in $(1-\cos k_{l})$ with the spatial direction index $1 \le l \le d$
where $d$ is the dimensionality.  In the small $|\vec{k}|$ (continuum
  limit),
$[1- \cos k_{l}] \to k_l^{2}/2$.  The particular case of a system with only finite range interactions that
exist up to a specified range $R$ on the lattice (the range being equal to 
a graph distance measuring the number of lattice steps beyond which
the interactions vanish) of the form of Eq.(\ref{PQ}) corresponds to $v(k) = P(z)$
with the order of the polynomial $M_{1}$ being equal to the interaction
range, $R= M_{1}$. Our result below includes such systems as
well as general systems with long range interactions.
For long range interactions
such as, e.g., the
screened Coulomb frustrated ferromagnet, $f(z)=1/(z+\lambda^2)$.
The considerations given below apply to the correlations
along any of the spatial directions $l$ (and as a particular
case, radially symmetric interactions for which the
correlations along all directions attain the 
same form). 

Returning to the form of Eq.(\ref{PQ}), the Fourier
space correlator of Eq.(\ref{sksq}) is given by 
\begin{eqnarray}
G(\vec{k}) = k_{B} T \frac{Q(z)}{F(z)}; ~~ F(z) = P(z) + \mu Q(z).
\label{gz}
\end{eqnarray}
On Fourier transforming Eq.(\ref{gz})
to real space to obtain the correlation
and modulation lengths, we see that
the zeros of $F(z)$ determine
these lengths. Expressed in terms
of its zeros, $F$ can be written as
\begin{eqnarray}
\label{Froots}
F(z) = A \prod_{j=1}^{M} (z- z_{j}),
\end{eqnarray}
where 
$M=\max[M_1,M_2]$.
Perusing Eq.(\ref{gz}), we see that 
$F$ is a polynomial in $z$ with 
real coefficients. As $F^*(z) = F(z^*)$
it follows that all roots of $F$
are either (a) real or (b) come in
complex conjugate pairs ($z_{j}= z_{i}^* \neq z_{i})$.
We now focus on the two cases separately.\\
(a) Real roots:
If a particular root $z_j=a^2$, $a\in\mathbb{R}$ then on
Fourier transforming Eq.(\ref{gz}) by the use of
the residue theorem, we obtain a term with a modulation length,
$L_D=2\pi/a$. Conversely,
if $z_j=-a^2$, we get a term with a correlation length,
$\xi=1/a$.\\
(b) Next we turn to the case of complex conjugate pairs
of roots.
If the pair of roots $z_j,z_{j}^*$ is not real, that is, $z_j=|z_{j}| e^{i\theta}$, then on Fourier transforming,
we obtain a term containing both a correlation length,
$\xi= (\sqrt{|z_{j}|}|\sin\frac{\theta}{2}|)^{-1}$, and modulation length,
$L_D=2\pi(\sqrt{|z_j|}|\cos\frac{\theta}{2}|)^{-1}$.\\ 
Putting all of the pieces together see that as 
(a) each real root of $F(z)$ contributes to either a correlation length or
a modulation length and (b) complex conjugate
pairs of non-real roots contribute to one correlation length and one
modulation length, {\em  the total number of correlation and
  modulation
lengths remains unchanged as the temperature ($\mu$) is varied.} The
total number of correlation + modulation lengths 
is given by the number of roots of $F(z)$ (that is, $M$). 
Thus, the system generally displays a net of $M$ correlation and
modulation lengths. 
This concludes our proof.
At very special temperatures, the Lagrange multiplier $\mu(T)$ may be
such that several poles degenerate into one -- thus lowering the number
of correlation/modulation lengths at those special temperatures. Also,
in case $M=M_2$, the total number of roots drops from $M_2$ to $M_1$
at $\mu=0$. What underlies multiple length-scales is the existence of terms of different ranges (different powers
of $z$ in the illustration above) -- not frustration. 

The same result can be proven using the transfer matrix method, for a
one-dimensional system with Ising spins. This is outlined in appendix \ref{TM}.
A trivial extension enables similar results for  other discrete 
spin systems (e.g., Potts spins). 

\subsection{First order transitions in the modulation 
length}
\label{1storder}
In this section, we show that there might be systems in which the
modulation length makes finite discontinuous jumps.
In these situations, the modulation length
does not diverge at a temperature $T^*$
(or set of such temperatures). The ground state modulation lengths
(the reciprocals of Fourier modes $\{\vec{q}_{i}\}$
minimizing the interaction
kernel) need not be 
continuous as a function
of the parameters that define the
interactions.  As we will simply illustrate below, 
in a manner that is mathematically 
similar to that appearing in the Ginzburg-Landau constructs, 
a ``first order transition''
in the value of the ground
state modulation lengths
can arise. Such a possibility is quite obvious and 
need not be expanded upon in depth.
As an illustrative example, let us 
consider the Range=3 interaction kernel
\begin{eqnarray}
v(k) =  a [\Delta+ \epsilon] + \frac{1}{2} b [\Delta+\epsilon]^{2} 
+ \frac{1}{3} c [\Delta + \epsilon]^{3}, 
\end{eqnarray}
\newline
with [$  0< \epsilon \ll 1$] and $c>0$.  
If  the parameters are such that $a >0$ and $b<0$, then $v(k)$ displays three
minima, i.e.  $[\Delta+\epsilon] =0$ and $[\Delta+\epsilon] = 
\pm m_{+}^{2}$ ,
where $m_{+}^{2} = \frac{1}{2c} [ -b+ \sqrt{b^{2}-4ac}]$.
the locus of points in the $ab$ plane where the three minima are equal
to one another
is defined by
$v(k)=0$. This leads to the relation
$m_{+}^{2} = -\frac{4a}{b}$.
Putting all of the pieces together, we see that $b= -4 \sqrt{ca/3}$
constitutes a line of ``first order transitions''. On traversing this line of "first order
transitions", the minimizing 
$[\Delta+ \epsilon]$ (and thus the minimizing wavenumbers)
changes discontinuously by 
$\Delta m = (- \frac{4a}{b})^{1/2} = (\frac{3a}{c})^{1/4}$.

\section{Example systems}
\label{appn}

In this section, we will investigate in detail
several frustrated systems. We will start our 
analysis by examining the screened Coulomb Frustrated 
Ferromagnet. A screened Coulomb interaction of screening length $\lambda^{-1}$
has the continuum Fourier
transformed interaction kernel $v(k) = [k^{2} +
\lambda^{2}]^{-1}$.
The lowest order non-vanishing derivative of $v_{L}(k)$ of order higher than two
is that of $p=3$.   
Invoking Eq.(\ref{selruleodd}), 
we find a modulation length 
that increases with increasing temperature as $T \to T_c^{+}$ 
(see also appendix \ref{explicit_form_p}, Eq.~(\ref{selrule3}) in particular) .

The dipolar interaction can be thought of as the $\delta\to0$ limit of the interaction,
\begin{eqnarray}
V_d=\frac{1}{[(\vec{x}-\vec{y})^2+\delta^2]^{3/2}}.
\end{eqnarray}
This form has a simple Fourier transform. 
In two spatial dimensions,
\begin{eqnarray}
v_d(k)=2\pi\delta^{-1}e^{-k\delta}.
\label{vdk2}
\end{eqnarray}
In three dimensions,
\begin{eqnarray}
v_d(k)=4\pi K_0(k\delta),
\label{vdk3}
\end{eqnarray}
$K_0$ being a modified Bessel function (see Eq.(\ref{K_0})).

In this case, we similarly find that the first non-vanishing derivative of $v_{L}$ is order
of order $p=3$ in the notation of Eq.(\ref{selruleodd}). This, as well as the detailed form
of Eq.(\ref{selrule3}) suggest an increasing modulation length with increasing temperature 
as $T \to T_c^{+}$.

\subsection{Numerical evaluation of the Correlation function}
\begin{figure*}[bt]
  \begin{center}
    A\epsfig{file=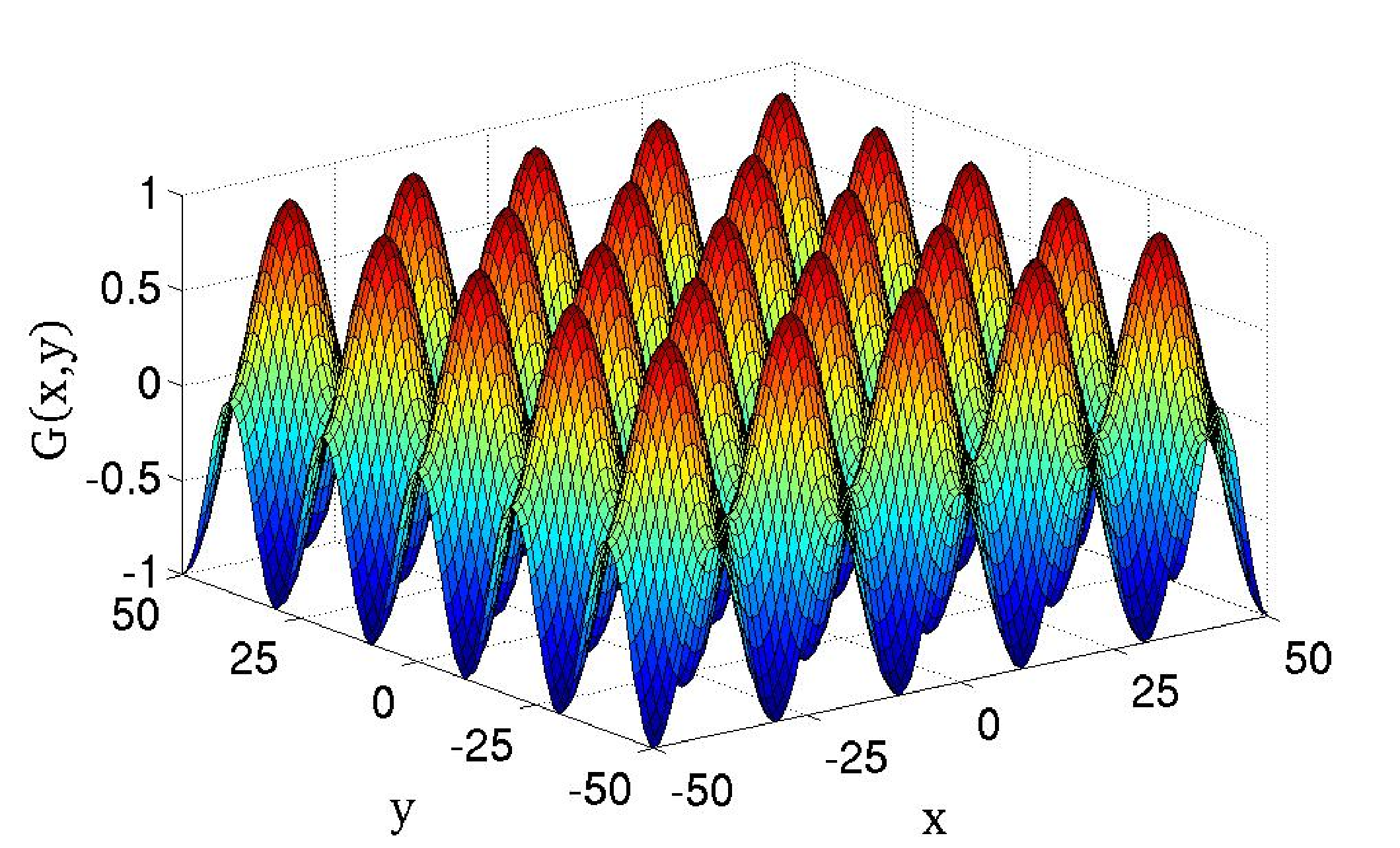,width=\columnwidth}
    \hfill
    B\epsfig{file=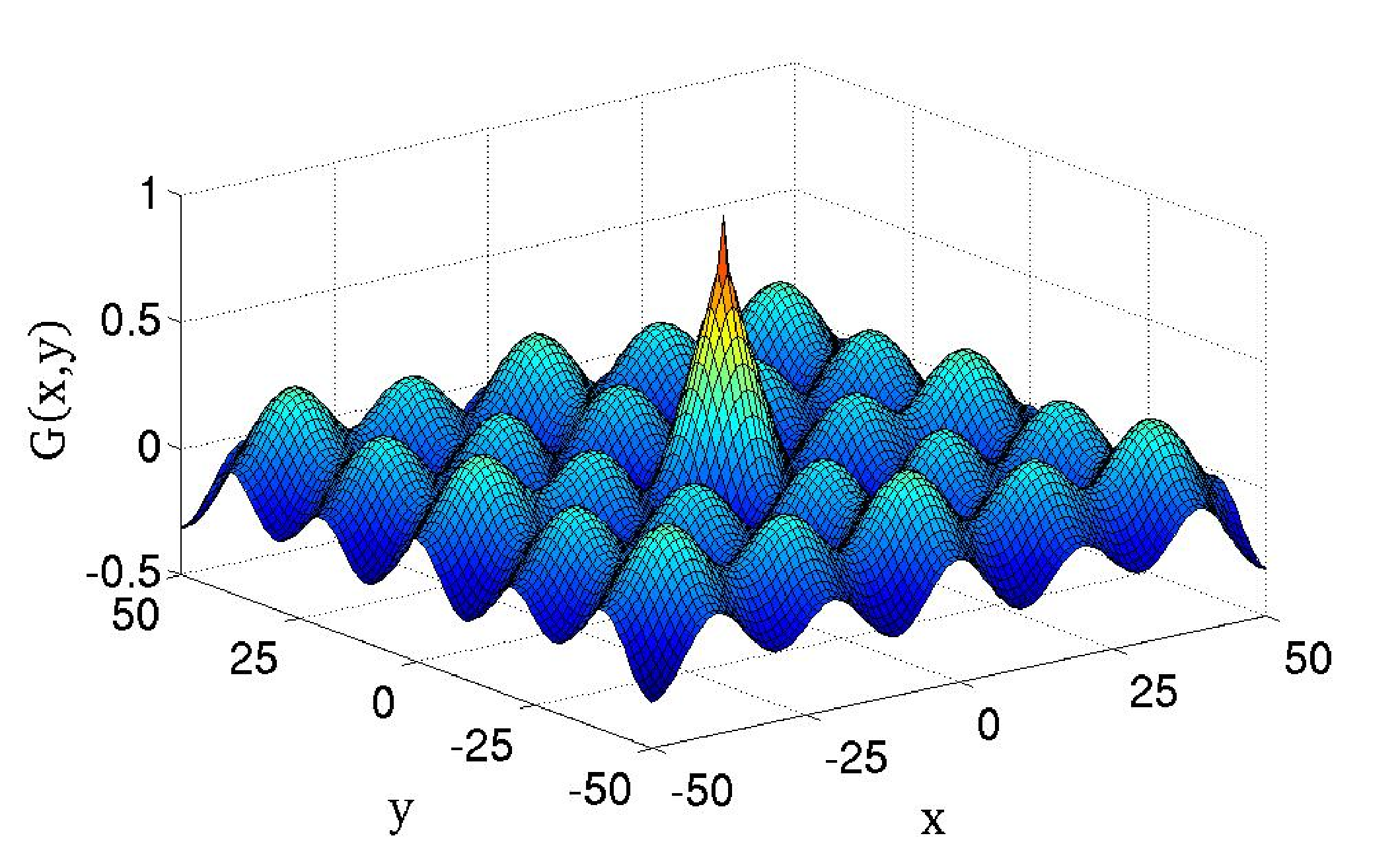,width=\columnwidth}\\
    C\epsfig{file=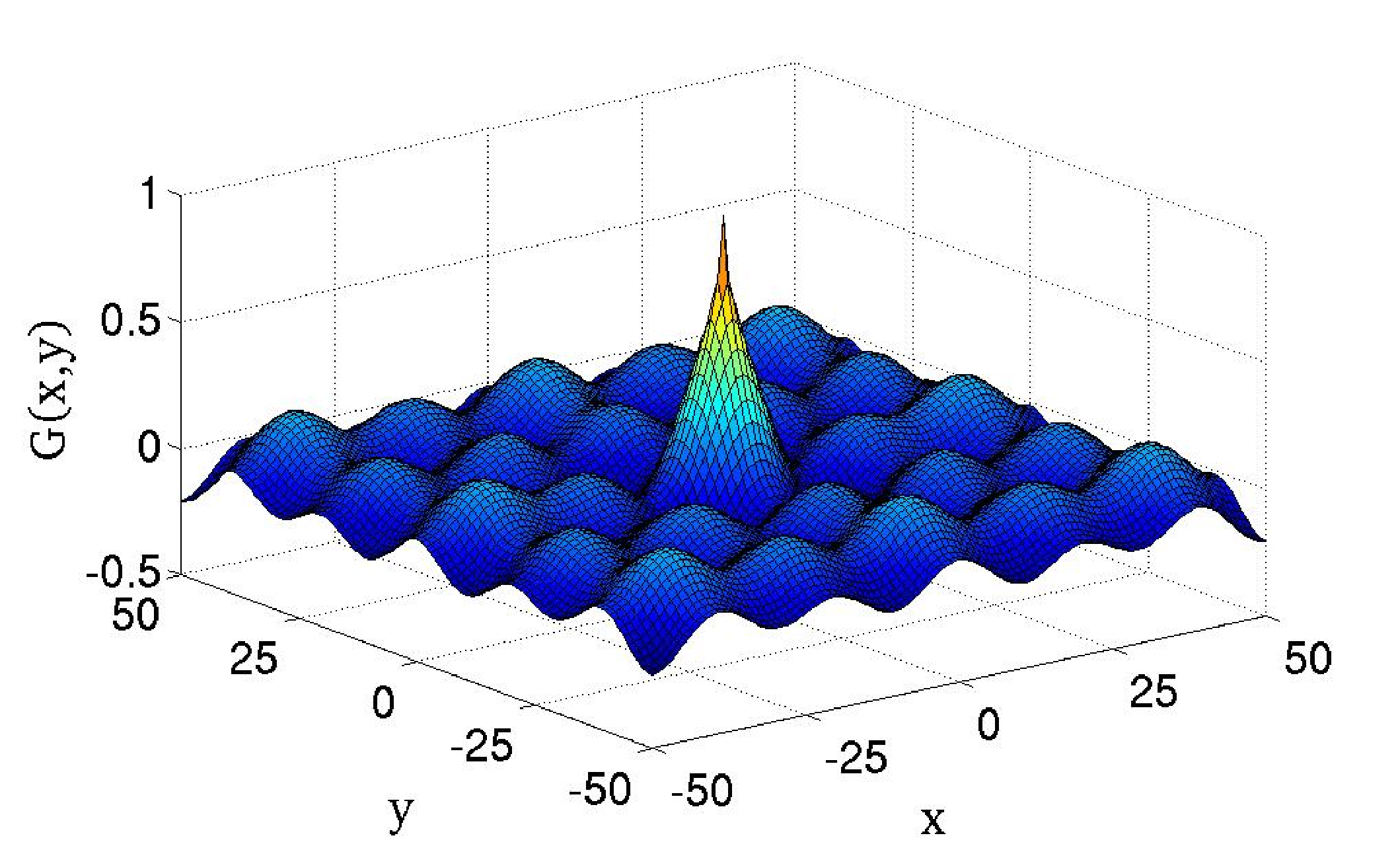,width=\columnwidth}
    \hfill
    D\epsfig{file=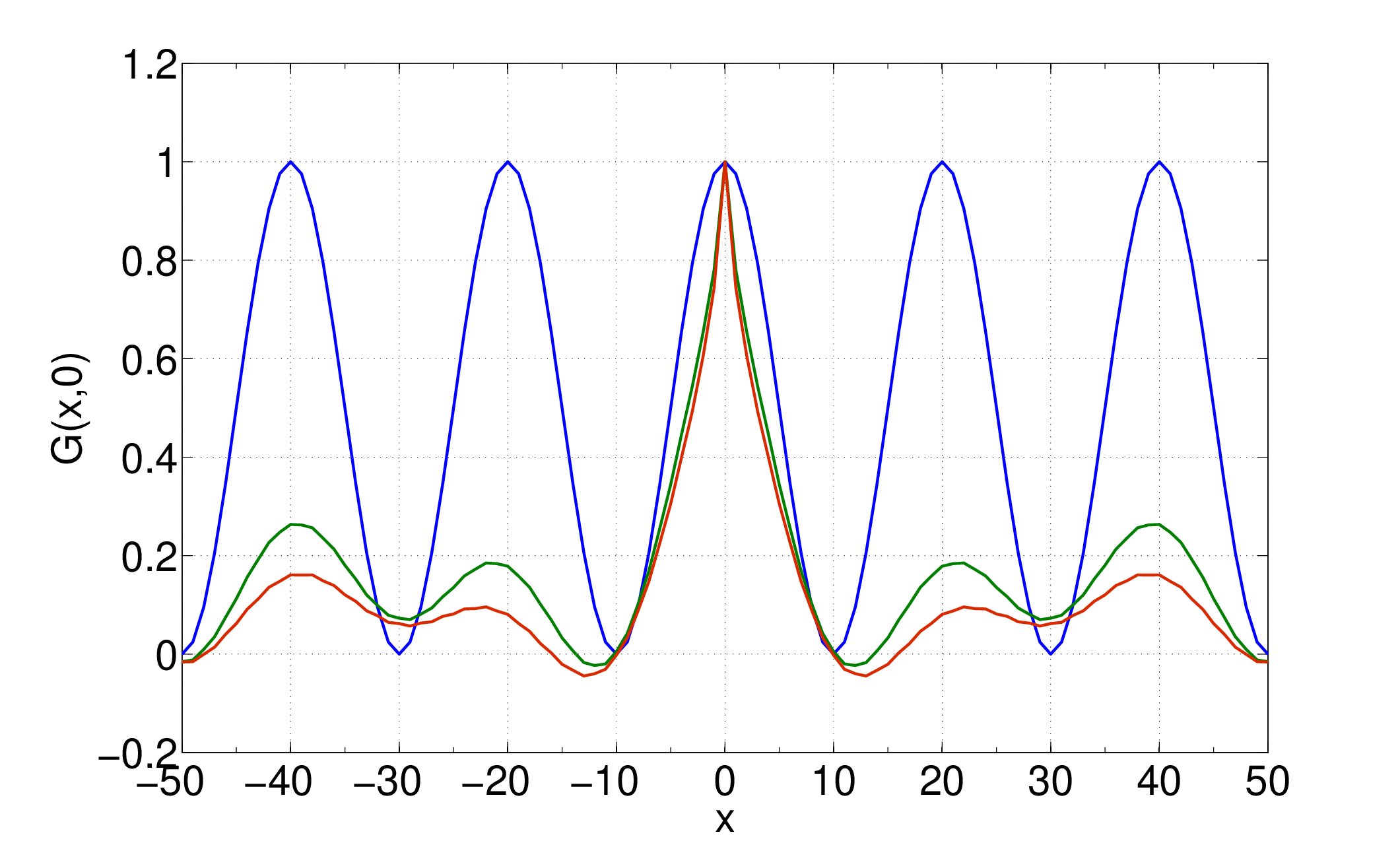,width=\columnwidth}
    \caption{(Color online.) The correlator $G(x,y)$ for a two dimensional Screened
      Coulomb Ferromagnet of size 100 by 100. $J=1$, $Q=0.0004$,
      Screening length$=100\sqrt{2}$. A: $\mu=\mu_{min}=-0.0874$, B:
      $\mu=\mu_{min}+0.001$, C: $\mu=\mu_{min}+0.003$, D: G(x,y) for
      $y=0$ for A(blue)[$L_D=20$], B(green)[$L_D=24$] and C(red)[$L_D=26$].}
    \label{fig:cou_100}
  \end{center}
\end{figure*}

\begin{figure*}[bt]
  \begin{center}
    A\epsfig{file=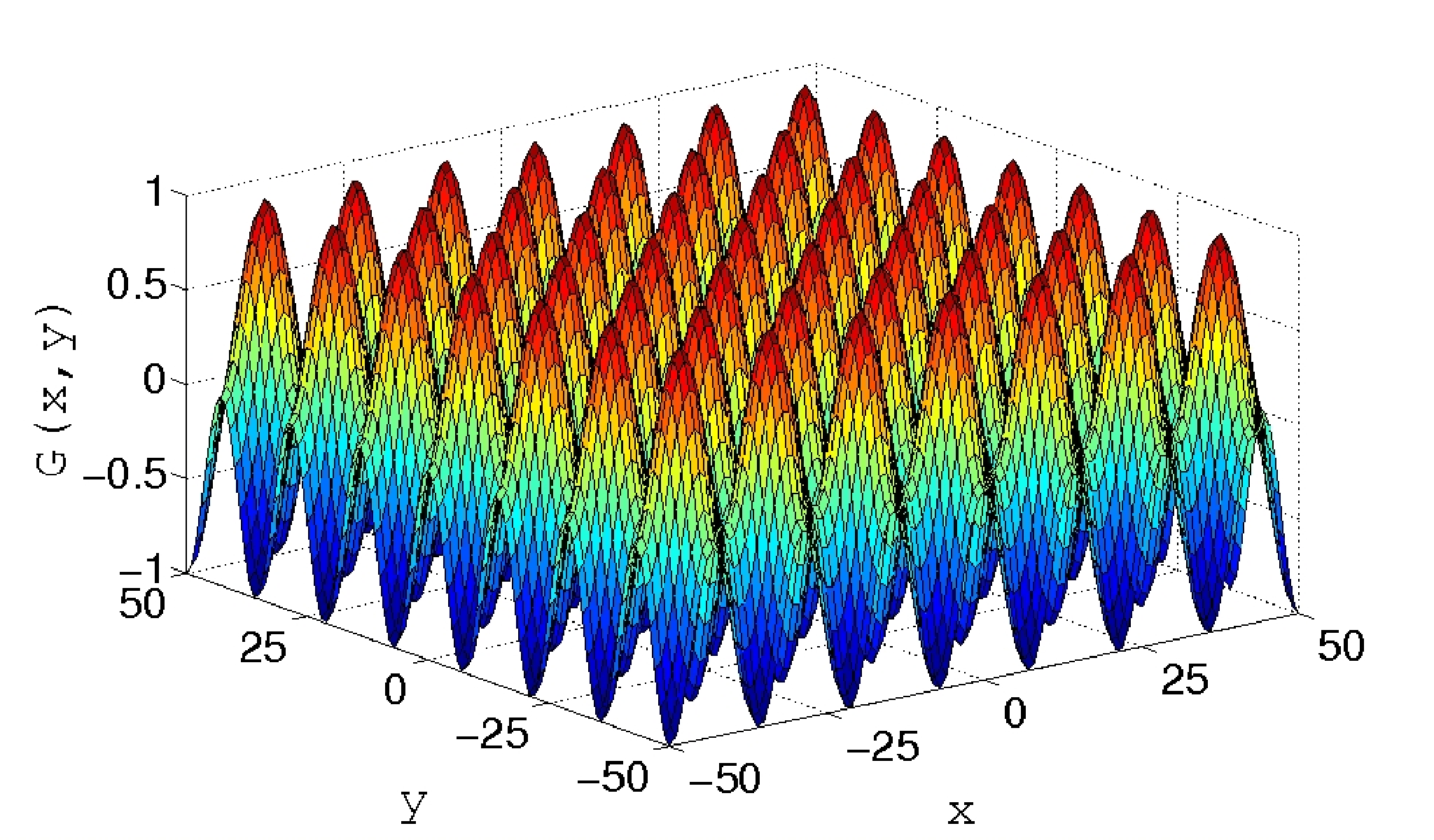,width=\columnwidth}
    \hfill
    B\epsfig{file=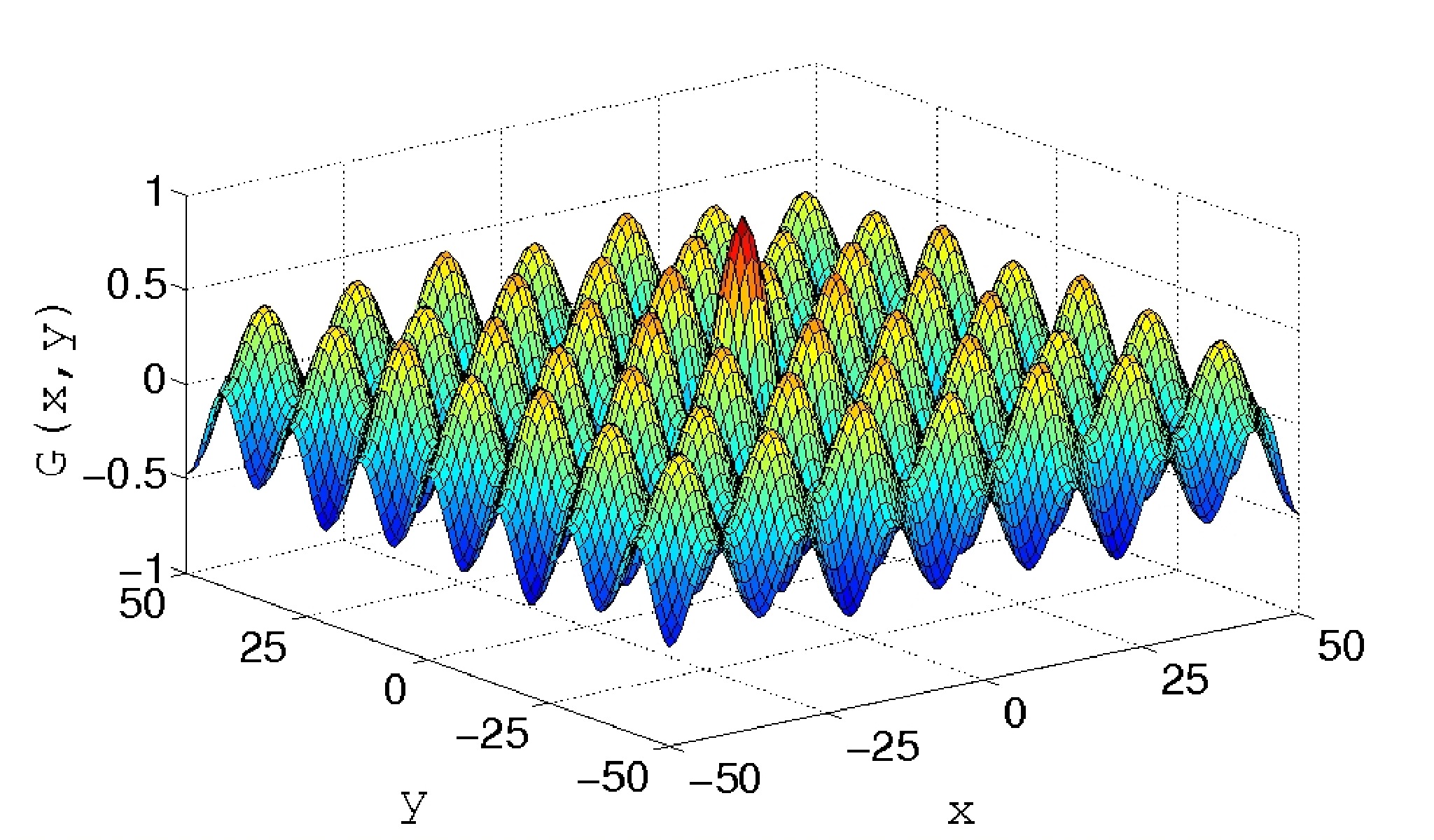,width=\columnwidth}\\
    C\epsfig{file=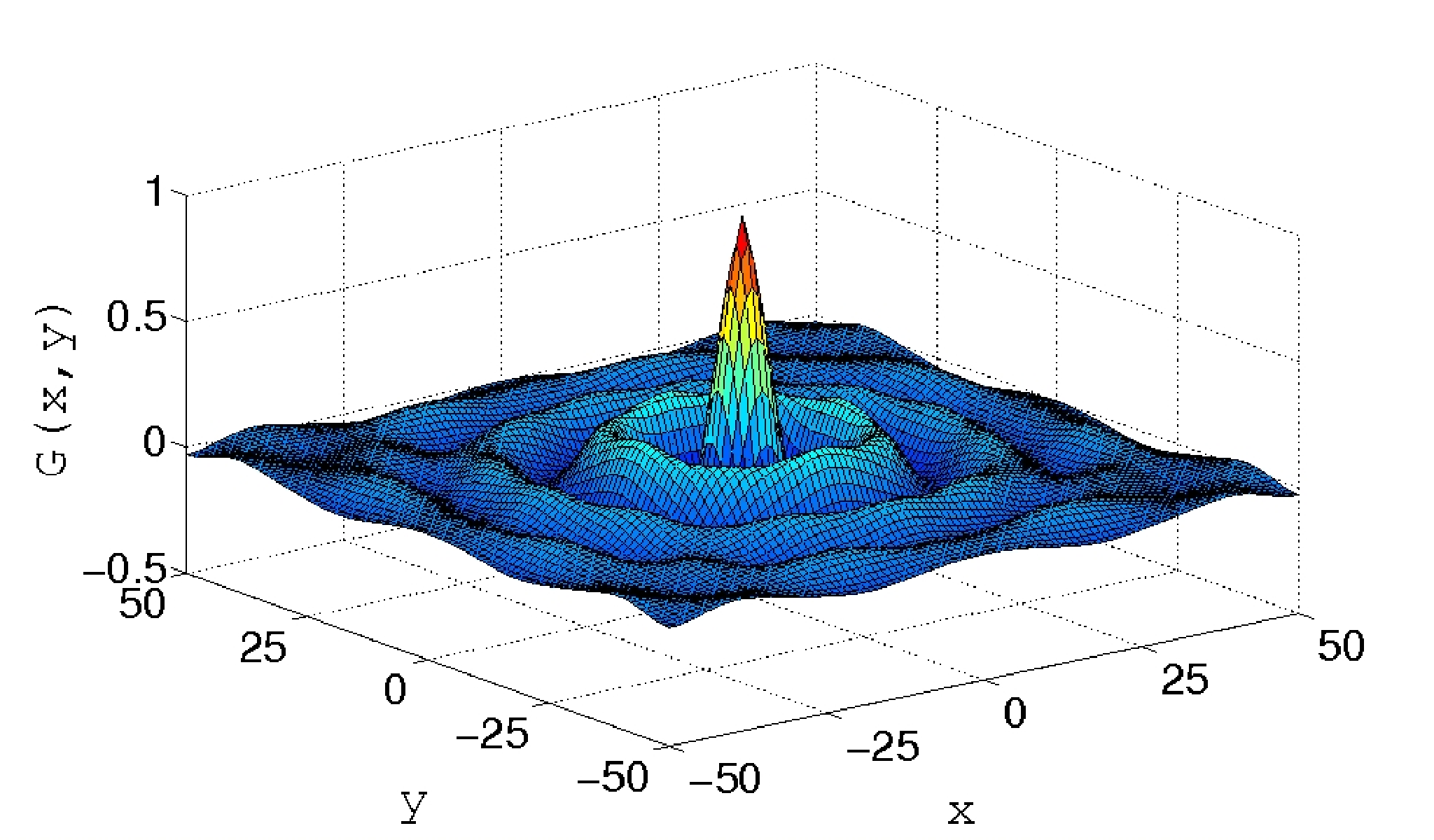,width=\columnwidth}
    \hfill
    D\epsfig{file=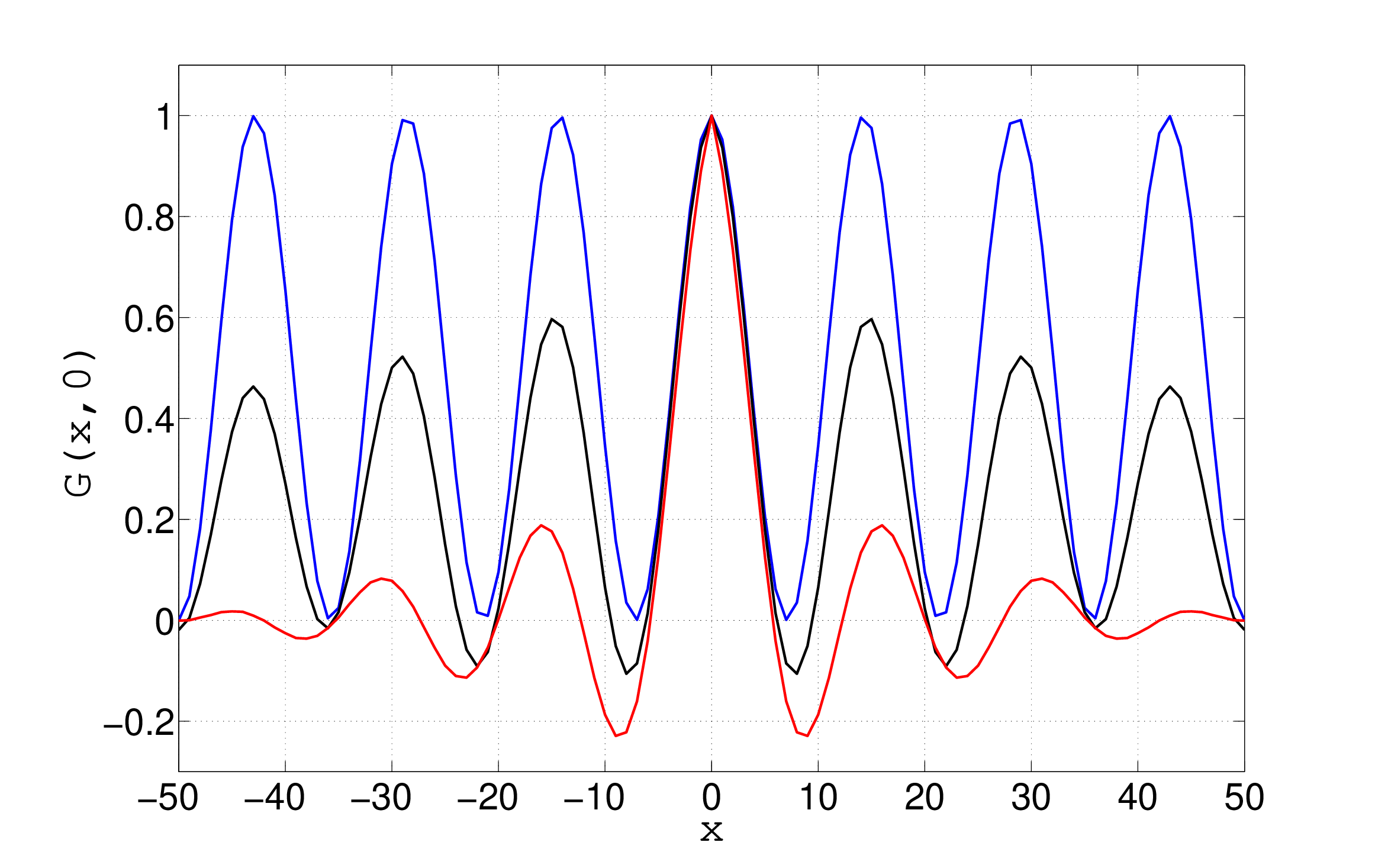,width=\columnwidth}
    \caption{(Color online.) The correlator $G(x,y)$ for a two dimensional Dipolar Ferromagnet of size 100 by 100. $J=1$, $Q=0.15$. A: $\mu=\mu_{min}=-1.1459$, B: $\mu=\mu_{min}+4\times 10^{-5}$, C: $\mu=\mu_{min}+1\times 10^{-3}$, D: G(x,y) for $y=0$ for A(blue)[$L_D=14$], B(green)[$L_D=15$] and C(red)[$L_D=16$].}
    \label{fig:dip_100}
  \end{center}
\end{figure*}

In Figs.(\ref{fig:cou_100},\ref{fig:dip_100}), we display a numerical evaluation of the
correlation function for the Coulomb frustrated ferromagnet and the
dipolar frustrated ferromagnet  (see Eqs.(\ref{Ham_JQ}, \ref{ccs}, \ref{dipint}))
on a two dimensional lattice of size
$100 \times 100$. 

\subsection{Coexisting short range and screened Coulomb interactions}
\label{Coul+}

In this section, we study the screened Coulomb frustrated ferromagnet
in more details.
In the continuum limit, the Fourier transform
of the interaction kernel of Eq.(\ref{Ham_JQ})
with $V_L(x)$ given by Eq.(\ref{ccs}) is
\begin{eqnarray}
v(k) = Jk^{2} + \frac{Q}{k^{2} + \lambda^{2}}.
\label{Yukawa+}
\end{eqnarray}
In appendix \ref{explainlong}, we provide explicit expressions for the dependence 
of $\mu$ on the temperature $T$. This dependence delineates the different
temperature regimes. 
For $T> T^{*}$ wherein
the temperature $T^{*}$ is set by 
\begin{eqnarray}
\mu(T^{*}) = J\lambda^{2}+ 2 \sqrt{JQ}, 
\label{mutst_cfif}
\end{eqnarray}
from Eq.(\ref{sksq}),
the pair correlator in $d=3$
dimensions is given by
\begin{eqnarray}
G(\vec{x}) = \frac{k_{B}T}{4 \pi J |\vec{x}|} \frac{1}{\beta^{2} - \alpha^{2}} 
\nonumber
\\ \times
[e^{- \alpha |\vec{x}|} (\lambda^{2} - \alpha^{2}) - e^{-\beta |\vec{x}|}
(\lambda^{2} - \beta^{2})].
\label{G1}
\end{eqnarray}
Here, 
\begin{eqnarray}
\alpha^{2}, \beta^{2} = 
\frac{\lambda^{2}+ \mu/J \mp \sqrt{(\lambda^{2} - \mu/J)^{2} - 4 Q/J}}{2}.
\label{3DHighT}
\end{eqnarray}
By contrast, for temperatures $T<T^{*}$, we obtain an analytic continuation of Eq.(\ref{G1})
to complex $\alpha$ and $\beta$,
\begin{eqnarray}
G(\vec{x}) = \frac{k_{B}T}{8 \alpha_{1} \alpha_{2} \pi J |\vec{x}|} 
e^{-\alpha_{1} |\vec{x}|} \nonumber
\\ \times
[(\lambda^{2} - \alpha_{1}^{2} + \alpha_{2}^{2}) \sin \alpha_{2}|\vec{x}|
+ 2 \alpha_{1} \alpha_{2} \cos \alpha_{2} |\vec{x}|],
\label{G2}
\end{eqnarray}
In Eq.(\ref{G2}), $\alpha = \alpha_{1} + i \alpha_{2} = \beta^{*}$. In a similar
fashion, in $d=2$ spatial dimensions, 
for $T>T^{*}$,
\begin{eqnarray}
G(\vec{x}) = \frac{k_{B}T}{2 \pi} \frac{1}{\beta^{2} - \alpha^{2}}
[(\lambda^{2}- \alpha^{2}) K_{0}(\alpha |\vec{x}|) \nonumber
\\ + (\beta^{2} - \lambda^{2})
K_{0}(\beta |\vec{x}|)].
\label{2DHighT}
\end{eqnarray}
As in the three dimensional case, the high temperature
correlator of Eq.(\ref{2DHighT}) may be analytically continued to lower temperatures, $T<T^{*}$,
for which $\alpha$ and $\beta$ become complex.

\subsubsection*{High temperature limit}

In the high temperature limit, in two spatial dimensions, from Eq.(\ref{2DHighT}), we have,
\begin{eqnarray}
\label{twodh}
G(\vec{x})=\frac{k_BT}{2\pi J}K_0\left(\sqrt{\frac{k_BT\Lambda^2}{4\pi
    J}}|\vec{x}|\right)
-\frac{8\pi}{k_BT\Lambda^4}K_0\left(\lambda |\vec{x}|\right).\nonumber\\
\end{eqnarray}
In three spatial dimensions, from Eq.(\ref{G1}), we have,
\begin{eqnarray}
\label{threedh}
G(\vec{x})=\frac{k_BT}{4\pi J|\vec{x}|}e^{-\sqrt{\frac{k_BT\Lambda^3}{6\pi^2J}}|\vec{x}|}-\frac{9\pi^3Q}{k_BT\Lambda^6|\vec{x}|}e^{-\lambda|\vec{x}|}.
\end{eqnarray}
In the unscreened case, in two spatial dimensions,
\begin{eqnarray}
G(\vec{x})&=&\frac{k_BT}{2\pi J}K_0\left(\sqrt{\frac{k_BT\Lambda^2}{4\pi
    J}}|\vec{x}|\right)\nonumber\\
&&-\frac{8\pi}{k_BT\Lambda^4}K_0\left(\sqrt{\frac{4\pi Q}{k_BT\Lambda^2}} |\vec{x}|\right).
\label{2dht}
\end{eqnarray}
In three spatial dimensions, 
\begin{eqnarray}
G(\vec{x}) &=& \Big[  \frac{k_BT}{4\pi J|\vec{x}|}e^{-\sqrt{\frac{k_BT\Lambda^3}{6\pi^2J}}|\vec{x}|}\nonumber
\\ &-&\frac{9\pi^3Q}{k_BT\Lambda^6|\vec{x}|}e^{-\sqrt{\frac{6\pi^2Q}{k_BT\Lambda^3}}|\vec{x}|} \Big].
\label{3dht}
\end{eqnarray}
From the above expressions, it is clear that the coefficients of the
terms corresponding to the diverging correlation length goes to zero
in the high temperature limit.

We note that two correlation lengths are manifest for all $(\mu-J\lambda^2)^{2}>4JQ$.
This includes all unfrustrated screened attractive 
Coulomb ferromagnets (those with $Q<0$)).
The evolution of the correlation functions may be traced by 
examining the dynamics of the poles in the complex $k$ plane
as a function of temperature. At high temperatures, 
correlations are borne
by poles that lie on the imaginary $k$ axis.

\subsubsection*{Thermal evolution of modulation length at low temperatures}

\begin{figure*}[thb]
  \begin{center}
    \includegraphics[width=2\columnwidth]{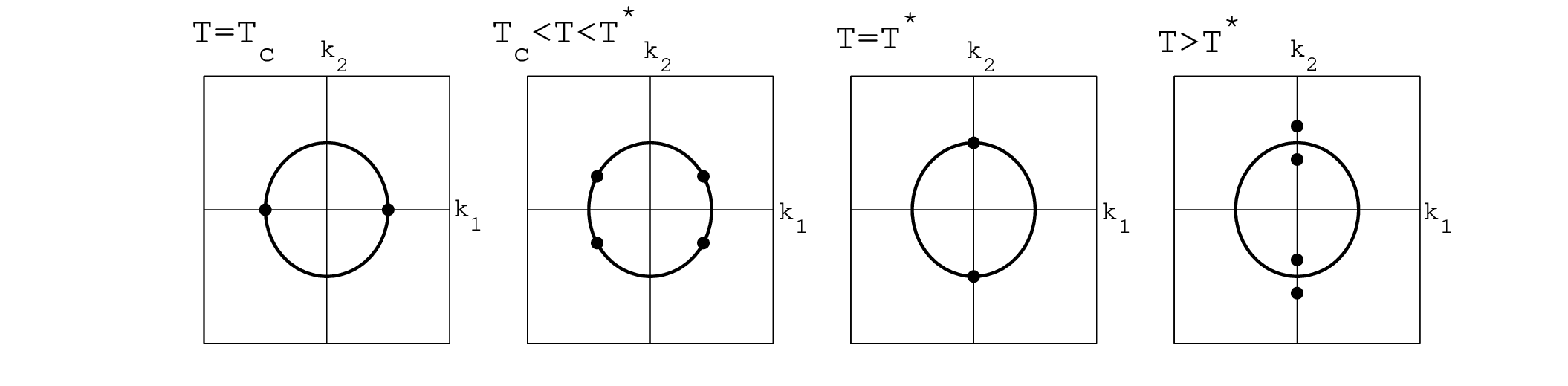}
    \caption{Location of the poles with increasing temperature (left
      to right) in the complex k-plane for the Coulomb frustrated
      ferromagnet. For temperatures below $T_c$,
      all the poles are real. Above $T_c$, the poles split in opposite
      directions of the real axis to give rise to two
      new complex poles. For $T_c<T<T^*$, we have complex
      poles. At $T^*$, pairs of such poles meet on the imaginary
      axis. Above $T^*$, the poles split along the imaginary
      axis. Thus, above $T^*$, the poles are purely imaginary.}
    \label{poles}
  \end{center}
\end{figure*}

\begin{figure*}[thb]
  \begin{center}
    \includegraphics[width=2\columnwidth]{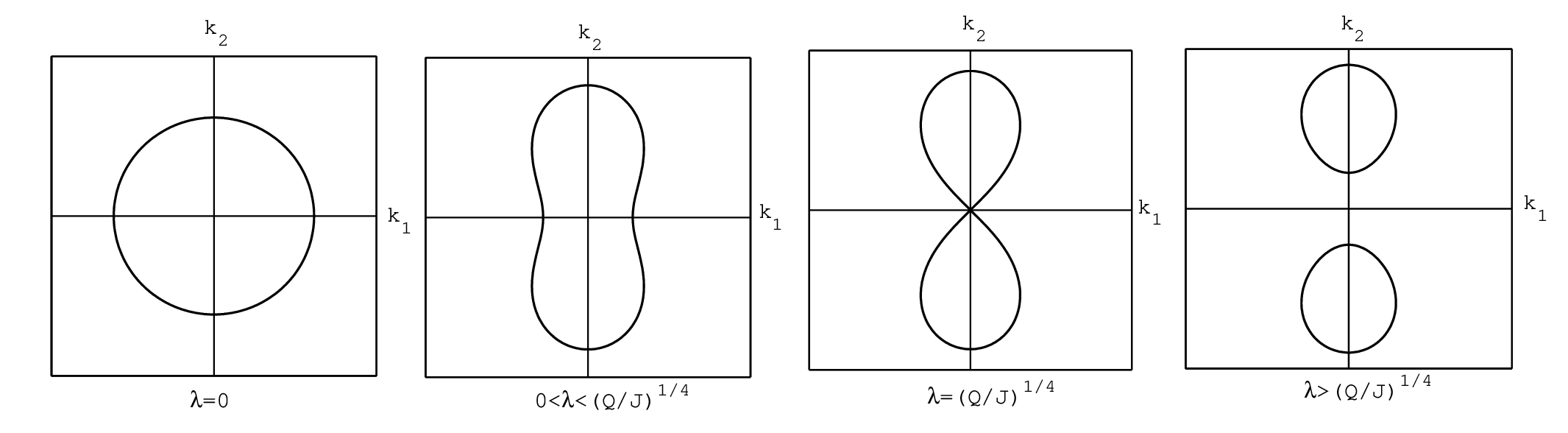}
    \caption{Trajectory of the poles in the complex $k$-plane for
      $T_c<T<T^*$ for the screened Coulomb ferromagnet. The screening
      length, $\lambda^{-1}$ decreases from left to right. In the first figure
      $\lambda=0$ and $\lambda>(Q/J)^{1/4}$ in the last figure.}
    \label{poles_scree}
  \end{center}
\end{figure*}

At $T=T^{*}$, the poles
merge in pairs at $k= \pm i \sqrt{\lambda^{2}+ \sqrt{Q/J}}$.
At lower temperatures,  $T< T^*$, the poles move off the imaginary axis
(leading in turn to oscillations in the correlation
functions). The norm of the poles, $|\alpha| = 
(Q/J+ \lambda^{2} \mu(T)/J)^{1/4}$ tends to a  constant
in the limit of vanishing screening ($\lambda^{-1} =0$)
wherein the after merging at $T=T^{*}$, the
poles slide along a circle [Fig. \ref{poles}]. In the low temperature
limit of the unscreened Coulomb ferromagnet,
the poles hit the real axis at finite $k$,
reflecting oscillatory modulations in the 
ground state. In the presence of 
screening, the pole trajectories are slightly
skewed [Fig. \ref{poles_scree}] yet for $Q/J> \lambda^{4}$,
$\alpha$ tends to the ground state
modulation wavenumber $\sqrt{\sqrt{Q/J} - \lambda^{2}}$. 
If the screening is sufficiently large,
i.e.,  if the screening length is shorter than 
the natural period favored by a balance between
the unscreened Coulomb interaction and the
nearest neighbor attraction ($\lambda > (Q/J)^{1/4}$),
then the correlation functions never exhibit 
oscillations. In such instances, the poles continuously
stay on the imaginary axis and, at low temperatures,
one pair of poles veers towards $k=0$ reflecting
the uniform ground state of the heavily screened
system.


\begin{figure}[hb]
  \begin{center}
    \includegraphics[width=\columnwidth]{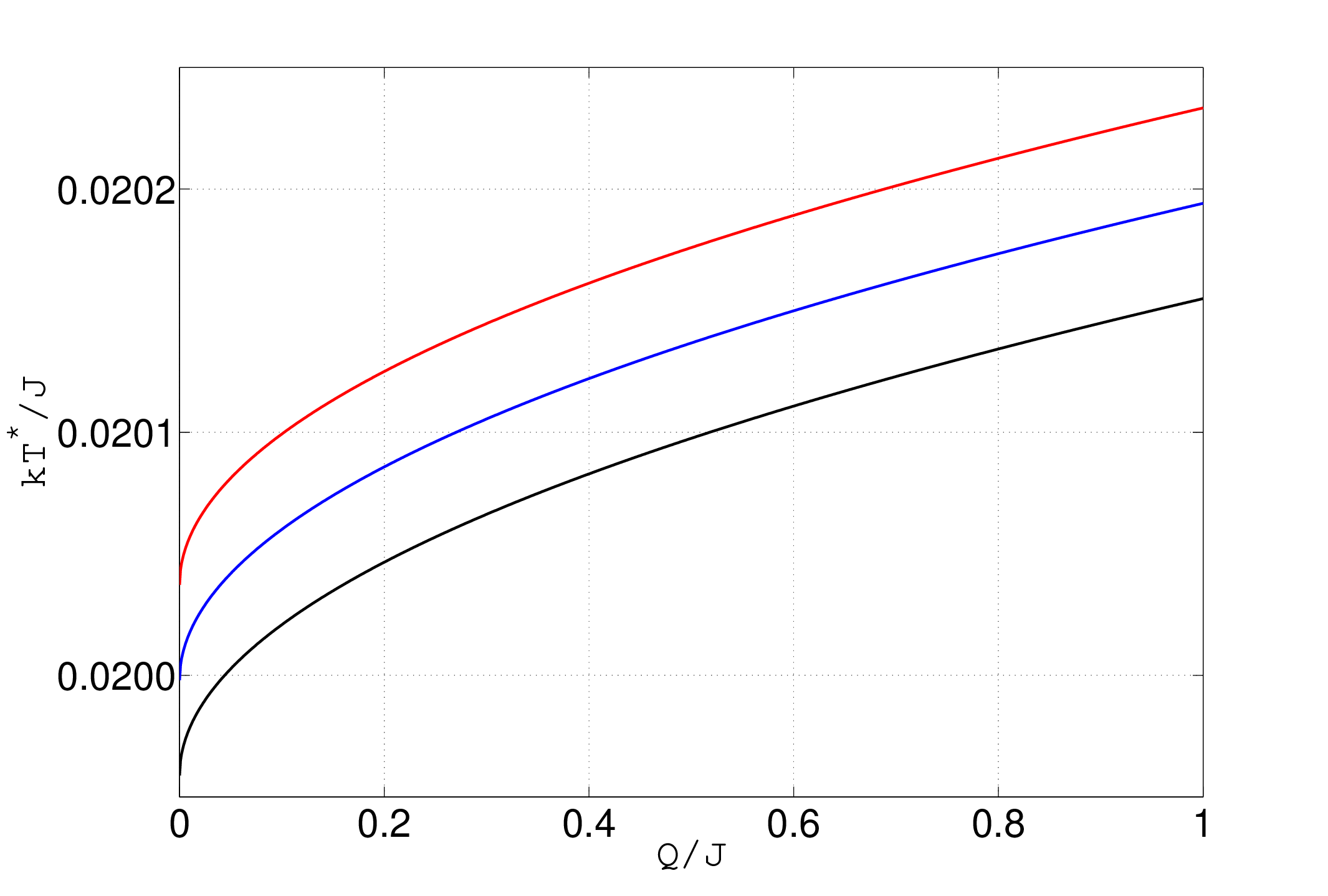}
    \caption{(Color online.) Temperature at which the modulation length diverges for a $100\times 100$ Coulomb frustrated ferromagnet plotted versus the relative strength of the Coulomb interaction with respect to the ferromagnetic interaction. [Blue:$\lambda=\lambda_0=1/(100\sqrt{2})$; Red:$\lambda=0.999\lambda_0$; Black:$\lambda=1.001\lambda_0$]}
    \label{Tstar}
  \end{center}
\end{figure}

To summarize, at high temperatures the pair correlator 
$G(x)$ is a sum of two decaying exponentials (one of 
which has a correlation length which diverges in 
the high temperature limit). For $T<T^{*}$ in under-screened
systems, one of the correlation lengths turns into
a modulation length characterizing low temperature oscillations.
At the cross-over temperature $T^{*}$, the modulation length
is infinite. As the temperature is progressively lowered,
the modulation length decreases in size -- until it
reaches its ground state value. 
The temperature $T^{*}(Q/J,\lambda)$ is a
``disorder line'' \cite{disorder1} like temperature [Fig \ref{Tstar}].
An analytical thermodynamic crossover does occur
at $T=T^{*}$. A large $n$ calculation illustrates that 
the internal energy per particle
\begin{eqnarray}
\frac{U}{N} = \frac{1}{2}(k_{B}T- \mu),
\label{Umu}
\end{eqnarray} 
To detect a crossover in $U$ and that in
other thermodynamic functions, the 
forms of $\mu$ both above and below $T^{*}$
may be derived from the spherical model normalization
condition to find that the real valued functional 
form of $\mu(T)$ changes [See appendix \ref{explainlong}]. 

The system starts to exhibit order
at the critical temperature $T=T_{c}$ given
by 
\begin{eqnarray}
\frac{1}{k_{B}T_{c}} = \int \frac{d^{d}k}{(2 \pi)^{d}} 
\frac{1}{v(\vec{k}) - v(\vec{q})}.
\label{tg}
\end{eqnarray}
For $Q/J > \lambda^{4}$, the modulus of the 
minimizing (ground state) wavenumber ($|\vec{q}|$) is
given by 
\begin{eqnarray}
q = \frac{2 \pi}{L_{D}^{g}} = \sqrt{\sqrt{Q/J} - \lambda^{2}},
\end{eqnarray}
with $L_{D}^{g}$ the ground state modulation length.
Associated with this wavenumber is the kernel
$v(\vec{q}) = 2 \sqrt{JQ} - J\lambda^{2}$
to be inserted in Eq.(\ref{tg}) for an evaluation of the
critical temperature $T_{c}$.
Similarly, the ground state wavenumber 
$\vec{q} =0$ whenever
$Q/J < \lambda^{4}$.
If $Q/J > \lambda^{4}$
and modulations transpire for temperatures $T<T^{*}$, the
critical temperature at which 
the chemical potential of Eq.(\ref{corr}),
$\mu(T_{c}) = J\lambda^{2} - 2 \sqrt{JQ}$,
is lower than the crossover temperature
$T^{*}$ (given by Eq.(\ref{mutst_cfif}))
at which modulations first start to appear.
The Screened Coulomb Ferromagnet
has
$T_{c}(Q/J=\lambda^{4})>0\mbox{ in $ d>4$  dimensions}$ and in any dimension
$T_{c}(Q/J>\lambda^{4})=0$. 
For small finite $n$, a first order Brazovskii transition
may replace the continuous transition occurring at $T_{c}$
within the large $n$ limit \cite{Braz}. Depending
on parameter values such an equilibrium transition may or may not
transpire before a possible glass transition may occur \cite{peter}. 
 
\subsubsection*{Domain length scaling in the Coulomb Frustrated Ferromagnet}
\label{exponent}
The characteristic length-scales are governed by the position of the poles of $[v(k)+\mu]^{-1}$. 
See Fig. \ref{poles} for an illustration
of the pole locations at low temperatures. 
For the frustrated Coulomb ferromagnet of Eq.(\ref{Yukawa+}) in the absence of screening ($\lambda^{-1} =0$), 
\begin{equation}
v(k)+\mu=\frac{J}{k^2}(k^4+\frac{\mu}{J}k^2+\frac{Q}{J}).
\label{domaincoul}
\end{equation}
Eq.(\ref{domaincoul}) enable us to determine, in our large $n$
analysis, the cross-over temperature $T^*$ at which
$\mu^*= \mu(T^*)=2\sqrt{JQ}$.
At $T=T^*$, the poles lie on the imaginary axis in $k$-plane.
As the temperature is lowered below $T^*$, the two poles bifurcate. This bifurcation gives 
rise to finite size spatial modulations. At temperatures $T< T^{*}$, the
four poles slide along a circle of fixed radius of size $(Q/J)^{1/4}$
(see Fig. \ref{poles}). At
zero temperature, these four poles merge in pairs to form two poles that lie on the real axis.
 The inverse modulation length is set 
by the absolute values of the real parts of the poles.
We will set $\mu \equiv (2\sqrt{JQ}-\delta\mu)$.
In the following, we will obtain the dependence of the real part of the poles on $\delta\mu$.
The poles of $1/(v(k) + \mu)$ are determined by 

\begin{eqnarray}
k_{pole}^{2}= - \frac{\mu}{2 J} \pm i \sqrt{\frac{Q}{J} - \frac{\mu^{2}}{4J^{2}}} = 
\sqrt{\frac{Q}{J}}e^{\pm 2 i\theta}.
\label{long}
\end{eqnarray}
At $\mu = \mu^*$, the angle $\theta = \pi/2$.
This point corresponds to the transition between (i) the 
high temperature region ($T> T^{*}$) 
wherein the system does not exhibit any modulations
and (ii) the low temperature region ($T<T^{*}$).
[See Fig. \ref{poles}.] Eq.(\ref{long}) implies that
$ \cos 2 \theta = [1 - \frac{\delta \mu}{\mu*}]$ 
or 
\begin{eqnarray} 
k_{pole,real}=\frac{\delta\mu^{1/2}}{2J^{1/2}}.
\end{eqnarray}
Thus, we get a crossover exponent of $1/2$.

\subsection{Full direction and location dependent dipole-dipole interactions}
In this subsection and the next, we consider systems where the spins are
three dimensional and the interactions have the appropriate directional dependence.
In this subsection, we will consider the effect of including the 
full dipolar interactions vis a vis the more commonly used scalar product form 
between two dipoles that is pertinent to two dimensional realizations. 
The dipolar interaction is given by
\begin{eqnarray}
H_{dip}=\sum_{\vec{x}\neq \vec{y}}\left[\frac{\vec{S}(\vec{x}) \cdot \vec{S}(\vec{y})}{r^3}-\frac{3[\vec{S}(\vec{x}) \cdot \vec{r}][\vec{S}(\vec{y}) \cdot \vec{r}]}{r^5}\right].
\label{hdip}
\end{eqnarray}
The two point correlator for a ferromagnetic system frustrated by this
interaction is given, in the large $n$ approximation, by
\begin{eqnarray}
G(\vec{x}) = k_{B} T
\int \frac{d^{d}k}{(2 \pi)^{d}} e^{i\vec{k} \cdot \vec{x}}
\left[\frac{2}{J\Delta(\vec{k})+Qv_d(k)+\mu}+\right.\nonumber\\
\left.\frac{1}{J\Delta(\vec{k})-2Qv_d(k)+\mu}\right],
\label{corr_fd}
\end{eqnarray}
where $v_d(k)$ is given by Eqs.(\ref{vdk2},\ref{vdk3}).
For temperatures $T \le T_{c}$, 
\begin{eqnarray}
\mu_{min}=-\min_{k\in\mathbb{R}} \{J\Delta(\vec{k})+Qv_d(k), \nonumber
\\ J\Delta(\vec{k})-2Qv_d(k) \}.
\end{eqnarray}
The Fourier transformed dipolar interaction kernel is positive definite, $v_d(k)>0$.
An unscreened dipolar interaction leads to a $v_{d}(k)$ that diverges
(tends to negative infinity) at its minimum at $k=0$. In the presence of both
upper and lower distance cutoffs (see, e.g., Eq.(\ref{dipint}) for a lower cutoff)
on the dipolar interaction,  the minimum of $v_{d}(k)$ attains a finite value
and the system has a finite critical temperature.

Examining Eq.(\ref{corr_fd}), we see that the introduction of the angular
dependence in the dipolar interaction 
changes the results that would be 
obtained if the angular dependence
were not included in a dramatic way.

{\bf{(i)}} New correlation and modulation lengths arise from the 
second term in Eq.(\ref{corr_fd}). 

{\bf{(ii)}} At low temperatures, the second term in Eq.(\ref{corr_fd})
becomes dominant as its poles have a smaller real part 
(and thus a larger correlation length) relative to the first 
term in Eq.(\ref{corr_fd}) that appears for an isotropic 
dipole-dipole interactions.

\subsection{Dzyaloshinsky- Moriya Interactions}
\label{DM}
As another example of a system with 
interactions having non-trivial directional dependence,  we consider a system of three
component spins with the Dzyaloshinsky-Moriya
interaction \cite{D-M} present along with the ferromagnetic interaction and a long
range interaction,
\begin{eqnarray}
H=-J\sum_{ \langle \vec{x},\vec{y} \rangle }\vec{S}(\vec{x}) \cdot \vec{S}(\vec{y})+\sum_{\langle \vec{x},\vec{y}\rangle} \vec{D} \cdot [\vec{S}(\vec{x})\times\vec{S}(\vec{y})]\nonumber\\
+Q\sum_{\vec{x} \neq \vec{y}}V_L(|\vec{x} - \vec{y}|)\vec{S}(\vec{x}) \cdot \vec{S}(\vec{y}).
\label{Ham_DM}
\end{eqnarray}
We diagonalize this interaction kernel to obtain a Hamiltonian of the
form,
\begin{eqnarray}
H=\sum_{\vec{x},\vec{y}}\sum_{a}\hat{S}^*_a(\vec{x})V_a(\vec{x},\vec{y})\hat{S}_a(\vec{y}).
\end{eqnarray}
The $\hat{S_a}$'s are linear combinations of the components of $\vec{S}$.
In a large $n$ approximation, the two point correlator is given by
\begin{eqnarray}
G(\vec{x}) = k_{B} T
\int \frac{d^{d}k}{(2 \pi)^{d}} e^{i\vec{k} \cdot \vec{x}}
\left[\frac{1}{J\Delta(\vec{k})+Qv_L(k)+\mu}+\right.\nonumber\\
\left.\frac{2(J\Delta(\vec{k})+Qv_L(k)+\mu)}{(J\Delta(\vec{k})+Qv_L(k)+\mu)^2+(D_1^2+D_2^2+D_3^2)[\Delta(\vec{k})]^2}\right].
\label{corr_DM}
\end{eqnarray}

The presence of the Dzyaloshinsky-Moriya interaction does not
alter the original poles and hence does not change the
original length-scales of the system. However, additional length-scales
arise due to the
second term in Eq.(\ref{corr_DM}).

A system of prominence where Dzyaloshinsky-Moriya interactions are important
is MnSi \cite{DMJ}.  The spiral order is naturally susceptible to glass like 
dynamics. \cite{new, DMJ}

\section{Conclusions}
\label{conc}
\begin{enumerate}
\item
We studied the evolution of the ground state modulation lengths
in frustrated Ising systems.
\item
We investigated, in large $n$ theories, the evolution of 
modulation and correlation 
lengths as a function of temperature
in different classes of systems.
\item
We proved that, in large $n$ theories, 
the combined sum of the number of correlation and the number of
modulation lengths is conserved.
We have also showed that there exists a diverging modulation length at
high temperatures for systems with long range interactions.
\item
We studied three dimensional dipolar systems. 
We found that the full dipolar interactions with angular dependence included,
changes the ground state of the system and also
adds new length-scales.
\end{enumerate}

It is a pleasure to acknowledge many discussions 
with Lincoln Chayes, 
Daniel Kivelson, Steven Kivelson, Michael Ogilvie,
Joseph Rudnick, Gilles Tarjus, and Peter
Wolynes. We also thank the Center for Materials Innovation --
Washington University in St. Louis for providing partial financial support for this project.

\begin{appendix}
\section{Transfer Matrix in the one-dimensional system with Ising spins}
\label{TM}
Thus far, we focused primarily on high dimensional continuous (large $n$) spin
systems. For completeness, we  review and illustrate how some similar conclusions
can be drawn for one dimensional Ising systems with finite ranged interactions
and briefly discuss trivial generalizations.
In particular, we show how the sum of the  number of modulation
and number of correlation lengths does not change as the 
temperature is varied. In Section(\ref{conser}), we illustrated
how this arises for general large $n$ systems.

For interactions of range $R$ in a one dimensional
Ising spin chain, the
transfer matrix, $T$ is of dimension $M={\cal{O}}(2^R)$. The correlation function
for large system size, takes the form
\begin{equation}
G(x)=\sum_{k=1}^{2^R-1} A_k \left(\frac{\lambda_k}{\lambda_0}\right)^x,
\end{equation}
where $\lambda_i$s are the eigenvalues of the transfer matrix.
Since the characteristic equation has real non-negative coefficients, from
Perron-Frobenius theorem, $\lambda_0$ is real, positive and is
non-degenerate.
The secular equation, $\det(T-\lambda I)=0$ is a polynomial in
$\lambda$ with real coefficients. Thus, two possibilities need to be
examined: real roots, and, complex conjugate pairs of roots.
Real eigenvalues $\lambda_p$ give terms with correlation length,
\begin{eqnarray}
\xi=\ln\left(\frac{\lambda_0}{|\lambda_p|}\right).
\end{eqnarray}
Complex conjugate eigenvalues, $\lambda_q$ and $\lambda_q^*$
correspond to the same correlation length and  modulation length,
given by,
\begin{eqnarray}
\xi=\ln\left(\frac{\lambda_0}{|\lambda_q|}\right),\\
L_D=\frac{2\pi}{\tan^{-1}\left(\frac{ \mbox{Im} \{ \lambda_q \}} {\mbox{Re} \{ \lambda_q \} }\right)}.
\end{eqnarray}

Thus, the total number of correlation and modulation lengths is
the order of the polynomial in $\lambda$ in the secular equation, or
simply the dimension of the transfer matrix --
${\cal{O}}(2^R)$. Similar to our conclusions for the high dimensional
continuous spin systems, this number does not vary with temperature.
For $q$ state Potts type spins, replicating the above arguments
mutatis mutandis, we find that the total number of
correlation and modulation lengths is ${\cal{O}}(q^{R})$.
Similarly, for such a system placed on a $d$ dimensional slab of finite 
extent in, at least, $(d-1)$ directions along which it
has a length of order ${\cal{O}}(l) > R$,
there will be ${\cal{O}}(q^{l^{d-1}})$ transfer 
matrix eigenvalues and thus an identical number
for the sum
of the number of modulation lengths
with the number of correlation lengths. 

The eigenvalues change from being complex below certain crossover
temperatures to being purely real above. These temperatures
form the ``disorder line''.

\section{Detailed expressions for $\delta L_D$ for different orders $p(\ge3)$ at which
the interaction kernel has its first non-vanishing derivative}
\label{explicit_form_p} 
If  the lowest order (larger than $p=2$) non-vanishing derivative of $v(k)$ at the minimizing wavenumber
$k_{0}$ (see Eq.(\ref{kmin})) is of order 
$p=3$ (the most common case) then, in the large $n$ limit, the change in the modulation 
length at temperatures $T>T_{c}$ ($\mu(T)=\mu_{min}+\delta\mu$)about its value at $T=T_{c}$ of Eq.(\ref{selruleodd}) is given by 
\begin{equation}
\delta L_D=-\frac{2\pi}{k_0^2}\frac{v^{(3)}(k_0)\delta\mu}{3(v^{(2)}(k_0))^2}.
\label{selrule3}
\end{equation}
We employ Eq.(\ref{selrule3}) in our analysis in Section(\ref{appn}).
If  the lowest order derivatives are of order $p=4$ or $5$ then, 
\begin{equation}
\delta L_D=\frac{2\pi}{k_0^2}\frac{v^{(5)}(k_0)(\delta\mu)^{2}}{30(v^{(2)}(k_0))^{3}}.
\end{equation}
Similarly, for $p=6$ or $7$,
\begin{equation}
\delta L_D=-\frac{2\pi}{k_0^2}\frac{v^{(7)}(k_0)(\delta\mu)^{3}}{630(v^{(2)}(k_0))^{4}},
\end{equation}
and so on.

\section{$\mu(T)$ for the screened Coulomb ferromagnet}
\label{explainlong}
We now briefly provide an explicit expression for the relation
between the large $n$ Lagrange multiplier $\mu$
and the temperature $T$ for the screened Coulomb ferromagnet. 
In three dimensions, with $\Lambda$ 
as an ultra-violet cutoff, 
at high temperatures [$T>T^{*}$], we get to the following
implicit equation for $\mu(T)$ in the case of
the screened Coulomb ferromagnet of Eq.(\ref{Yukawa+}),
\\
\begin{eqnarray}
&&\frac{1}{T} = \frac{\Lambda}{2 \pi^{2}} +
\frac{\sqrt{2}}{4 \pi^{2} p}\nonumber
\\ &&\times \Big( \frac{\lambda^{2} \mu - \mu^{2} 
+ \mu p - 2Q}{\sqrt{\lambda^{2}+ \mu + p}} \tan^{-1} 
(\frac{\Lambda \sqrt{2}}{\sqrt{\lambda^{2} + \mu + p}}) \nonumber
\\ && - \frac{\lambda^{2} \mu - \mu^{2} + \mu p + 2Q}
{\sqrt{\lambda^{2} + \mu - p}}
\tan^{-1}(\frac{\Lambda \sqrt{2}}{\sqrt{\lambda^{2}  + \mu - p}}) \Big).
\label{explicitT}
\end{eqnarray}
In Eq.(\ref{explicitT}), we employed the shorthand 
$p \equiv \sqrt{(\mu - \lambda^{2})^{2} - 4Q}$.
The parameter $p$ 
vanishes at the crossover temperature $T^{*}$ at which 
a divergent modulation length makes an appearance, $p(T=T^{*})=0$.
At low temperatures, $T<T^{*}$, $p$ becomes imaginary and 
an analytical crossover occurs to another real functional form. 

\section{Proof that $\mu(T)$ is an analytic function of $T$}
\label{mutanalytic}
In this appendix, we illustrate that in the large $n$ limit, the thermodynamic functions are analytic for all
temperatures $T>T_{c}$, including the discussed cross-over
temperature $T=T^*$ (hence justifying the use of the term "cross-over") 
of, e.g, the Coulomb frustrated ferromagnet of Eqs.(\ref{Ham}, \ref{Yukawa+}).

From Eq. (\ref{g1mut}), using,
\begin{eqnarray}
\mu > \mu_{\min}=-\min{v(k)},
\end{eqnarray}
it is clear that $\mu(T)$ is a continuous
function of $T$. Differentiating, 
\begin{eqnarray}
\frac{d\mu}{d(k_BT)}=\left[(k_BT)^2\int \frac{d^dk}{(v(k)+\mu)^2}\right]^{-1},
\end{eqnarray}
and,
\begin{eqnarray}
\frac{d^2\mu}{d(k_BT)^2}=-\frac{2}{k_BT}\frac{d\mu}{d(k_BT)}+\nonumber\\
2(k_BT)^2\left(\frac{d\mu}{d(k_BT)}\right)^3\int \frac{d^dk}{(v(k)+\mu)^3}
\end{eqnarray}
with the integrations performed over the first Brillouin zone on the lattice 
(or up to some ultra-violet cutoff $\Lambda$ in the continuum). 
The first two derivatives are thus always finite so long as the
integration range is finite.
All higher order derivatives are sum of terms which are products of
lower order derivatives, $(k_BT)^a$ and
$\int \frac{d^dk}{(v(k)+\mu)^b}$, where $a$ and $b$ are
  integers, with $b>0$. Thus, for finite integration range, $\mu(T)$
  is an analytic function of $T$. In the large $n$ limit, the internal energy per site,
  $U/N = [k_BT- \mu(T)]/2$. Our result concerning the analyticity of $\mu(T)$
  implies that the internal energy is analytic and thus all of its derivatives
  and all other thermodynamic potentials. 

\end{appendix}

\bibliography{mybiblio}

\begin{thebibliography}{38}
\expandafter\ifx\csname natexlab\endcsname\relax\def\natexlab#1{#1}\fi
\expandafter\ifx\csname bibnamefont\endcsname\relax
  \def\bibnamefont#1{#1}\fi
\expandafter\ifx\csname bibfnamefont\endcsname\relax
  \def\bibfnamefont#1{#1}\fi
\expandafter\ifx\csname citenamefont\endcsname\relax
  \def\citenamefont#1{#1}\fi
\expandafter\ifx\csname url\endcsname\relax
  \def\url#1{\texttt{#1}}\fi
\expandafter\ifx\csname urlprefix\endcsname\relax\def\urlprefix{URL }\fi
\providecommand{\bibinfo}[2]{#2}
\providecommand{\eprint}[2][]{\url{#2}}

\bibitem[{isi()}]{ising}
\bibinfo{note}{E. Ising, Zeits. f. Physik. {\bf 31}, 253 (1925).}

\bibitem[{lon()}]{long}
\bibinfo{note}{T. Dauxios, S. Ruffo, E. Arimondo, M. Wilkens (Eds.), ``Dynamics
  and Thermodynamics of Systems with Long Range Interactions'', Lecture Notes
  in Physics {\bf 602}, Springer (2002).}

\bibitem[{lie()}]{lieb}
\bibinfo{note}{A. Giuliani, J. L. Lebowitz and E. H. Lieb, Phys. Rev. B {\bf
  76}, 184426 (2007); A. Giuliani, J. L. Lebowitz and E. H. Lieb, Phys. Rev. B
  {\bf 74}, 064420 (2006).}

\bibitem[{vin()}]{vindigni}
\bibinfo{note}{A. Vindigni, N. Saratz, O. Portmann, D. Pescia, and P. Politi,
  Phys. Rev. B {\bf 77}, 092414 (2008).}

\bibitem[{ort()}]{ortix}
\bibinfo{note}{C. Ortix, J. Lorenzana, and C. Di Castro, Phys. Rev. B {\bf 73},
  245117 (2006).}

\bibitem[{gul()}]{gulacsi}
\bibinfo{note}{I. Daruka and Z. Gul\'acsi, Phys. Rev. E {\bf 58}, 5403 (1998).}

\bibitem[{bar({\natexlab{a}})}]{barci}
\bibinfo{note}{D. G. Barci and D. A. Stariolo, Phys. Rev. B {\bf 79}, 075437
  (2009).}

\bibitem[{Fog()}]{Fogler}
\bibinfo{note}{M. M. Fogler, arXiv:cond-mat/0111001, p. 98-138, in High
  Magnetic Fields: Applications in Condensed Matter Physics and Spectroscopy,
  ed. by C. Berthier, L.-P. Levy, G. Martinez (Springer-Verlag, Berlin, 2002).}

\bibitem[{amp()}]{amp}
\bibinfo{note}{Hyung-June Woo, C. Carraro, D. Chandler, Phys. Rev. E {\bf 52},
  6497 (1995); F. Stilinger, J. Chem. Phys. {\bf 78}, 4655 (1983); L. Leibler,
  Macromolecules {\bf 13}, 1602 (1980); T. Ohta and K. Kawasaki, Macromolecules
  {\bf 19}, 2621 (1986).}

\bibitem[{vor({\natexlab{a}})}]{vortex3}
\bibinfo{note}{H. Kleinert, ``Gauge Fields in Condensed Matter'', World
  Scientific (1989), volume II .}

\bibitem[{vor({\natexlab{b}})}]{vortex1}
\bibinfo{note}{P. H. Chavanis, ``Statistical Mechanics of Two Dimensional
  Vortices and Stellar systems'' in Ref.[\cite{long}].}

\bibitem[{vor({\natexlab{c}})}]{vortex2}
\bibinfo{note}{H. Kleinert, ``Gauge Fields in Condensed Matter'', World
  Scientific (1989), volume I .}

\bibitem[{Seu()}]{Seul}
\bibinfo{note}{M. Seul and D. Andelman, Science {\bf 267}, 476 (1995).}

\bibitem[{wav()}]{wavpart}
\bibinfo{note}{Y. Elskens, ``Kinetic Theory for Plasmas and Wave-particle
  Hamiltonian Dynamics'', in \cite{long}; Y. Elskens and D. Escande,
  ``Microscopic Dynamics of Plasmas and Chaos'', IOP publishing, Bristol
  (2002).}

\bibitem[{ste()}]{steve}
\bibinfo{note}{V. J. Emery and S. A. Kivelson, Physica C {\bf 209}, 597
  (1993).}

\bibitem[{us()}]{us}
\bibinfo{note}{L. Chayes, V. J. Emery, S. A. Kivelson, Z. Nussinov, and G.
  Tarjus, Physica A {\bf 225}, 129 (1996).}

\bibitem[{zoh({\natexlab{a}})}]{zohar}
\bibinfo{note}{Z. Nussinov , J. Rudnick, S. A. Kivelson, and L. N. Chayes,
  Phys. Rev. Letters 83, 472 (1999).}

\bibitem[{\citenamefont{L\"ow et~al.}(1994)\citenamefont{L\"ow, Emery,
  Fabricius, and Kivelson}}]{Low}
\bibinfo{author}{\bibfnamefont{U.}~\bibnamefont{L\"ow}},
  \bibinfo{author}{\bibfnamefont{V.~J.} \bibnamefont{Emery}},
  \bibinfo{author}{\bibfnamefont{K.}~\bibnamefont{Fabricius}},
  \bibnamefont{and} \bibinfo{author}{\bibfnamefont{S.~A.}
  \bibnamefont{Kivelson}}, \bibinfo{journal}{Phys. Rev. Lett.}
  \textbf{\bibinfo{volume}{72}}, \bibinfo{pages}{1918} (\bibinfo{year}{1994}).


\bibitem[{car()}]{carlson}
\bibinfo{note}{E. W. Carlson, V. J. Emery, S. A. Kivelson, D. Orgad, ``Concepts
  in High Temperature Superconductivity'' in ``The Physics of Superconductors''
  ed. K.H. Bennemann and J.B. Ketterson (Springer-Verlag 2004), 180 pages. See
  also arXiv:cond-mat/0206217.}

\bibitem[{nas()}]{nasci}
\bibinfo{note}{V.B. Nascimento et. al., arXiv:0905.3194.}

\bibitem[{che()}]{cheong}
\bibinfo{note}{S- W. Cheong et al., Phys. Rev. Lett. {\bf 67}, 1791 (1991).}

\bibitem[{gol()}]{golosov}
\bibinfo{note}{D. I. Golosov, Phys. Rev. B, vol. 67, 064404 (2003).
  (arXiv:cond-mat/0206257).}

\bibitem[{\citenamefont{Kivelson et~al.}(1995)\citenamefont{Kivelson, Kivelson,
  Zhao, Nussinov, and Tarjus}}]{dk}
\bibinfo{author}{\bibfnamefont{D.}~\bibnamefont{Kivelson}},
  \bibinfo{author}{\bibfnamefont{S.~A.} \bibnamefont{Kivelson}},
  \bibinfo{author}{\bibfnamefont{X.}~\bibnamefont{Zhao}},
  \bibinfo{author}{\bibfnamefont{Z.}~\bibnamefont{Nussinov}}, \bibnamefont{and}
  \bibinfo{author}{\bibfnamefont{G.}~\bibnamefont{Tarjus}},
  \bibinfo{journal}{Physica A: Statistical and Theoretical Physics}
  \textbf{\bibinfo{volume}{219}}, \bibinfo{pages}{27 } (\bibinfo{year}{1995}),
  ISSN \bibinfo{issn}{0378-4371}.

\bibitem[{pet()}]{peter}
\bibinfo{note}{J. Schmalian and P. G. Wolynes, Phys. Rev. Lett. {\bf 85}, 836
  (2000); H. Westfahl, Jr., J. Schmalian, and P. G. Wolynes, Phys. Rev. B {\bf
  64}, 174203 (2001).}

\bibitem[{Gil()}]{Gilles}
\bibinfo{note}{G. Tarjus, S. A. Kivelson, Z. Nussinov and P. Viot, J. Phys
  Cond. Matt. {\bf 17}, 50 (2005).}

\bibitem[{new()}]{new}
\bibinfo{note}{Z. Nussinov, Phys. Rev. B {\bf 69}, 014208 (2004).}

\bibitem[{rev()}]{reviewGilles}
\bibinfo{note}{G. Tarjus, S. A. Kivelson, Z. Nussinov, and P. Viot, J. Phys:
  Condens. Matter {\bf17}, R1143 (2005).}

\bibitem[{der()}]{der}
\bibinfo{note}{B. V. Derjaguin and L. Landau, Acta Physiochim, URSS {\bf 14},
  633 (1941); E. J. Verwey and J. T. G. Overbeek {\em Theory of Stability of
  Lyophobic Colloids} (Elsevier, Amsterdam, 1948).}

\bibitem[{rei()}]{reich}
\bibinfo{note}{C. Reichhardt and C. J. Olson, Phys. Rev. Lett. {\bf 88}, 248301
  (2002).}

\bibitem[{bar({\natexlab{b}})}]{barre}
\bibinfo{note}{J. Barre, D. Mukamel, S. Ruffo, ``Ensemble inequivalence in mean
  field models of magnetism'' in \cite{long}.}

\bibitem[{azb()}]{azbel}
\bibinfo{note}{Mark Ya. Azbel Phys. Rev. E {\bf 68}, 050901 (2003).}

\bibitem[{zoh({\natexlab{b}})}]{zohar_com}
\bibinfo{note}{Z. Nussinov, arXiv:cond-mat/0105253 (2001) -- in particular, see
  footnote [20] therein for the Ising ground states.}

\bibitem[{kac()}]{kac}
\bibinfo{note}{T. H. Berlin and M. Kac, Phys. Rev. {\bf 86}, 821 (1952)}.

\bibitem[{\citenamefont{Stanley}(1968)}]{stanley}
\bibinfo{author}{\bibfnamefont{H.~E.} \bibnamefont{Stanley}},
  \bibinfo{journal}{Phys. Rev.} \textbf{\bibinfo{volume}{176}},
  \bibinfo{pages}{718} (\bibinfo{year}{1968}).

\bibitem[{dis()}]{disorder1}
\bibinfo{note}{J. Stephenson, Phys. Rev B, {\bf 1}, 4405 (1970); J. Stephenson,
  Can. J. Phys, {\bf 48}, 1724 (1970); N. Alves Jr. and C. S. O. Yokoi, Braz.
  J. Phys., {\bf 30}, 4 (2000); W. Selke, Phys. Rep., {\bf 170}, 213 (1988).}

\bibitem[{Bra()}]{Braz}
\bibinfo{note}{S. Brazovskii, Sov. Phys. JETP {\bf 41}, 85 (1975).}

\bibitem[{D-M()}]{D-M}
\bibinfo{note}{I. Dzyaloshinsky, J. Phys. Chem. Solids {\bf 4}, 241 (1958); T.
  Moriya, Phys. Rev {\bf 120}, 1, 91 (1960).}

\bibitem[{DMJ()}]{DMJ}
\bibinfo{note}{J. Schmalian and M. Turlakov, Phys. Rev. Lett. {\bf{93}}, 036405
  (2004).}

\end{thebibliography}

\end{document}